\documentclass[nobibnotes,amsmath,amssymb,notitlepage,twocolumn,prb]{revtex4-2}

\usepackage{graphicx}
\usepackage{bm,braket}
\usepackage{comment}
\usepackage{dcolumn}
\usepackage{ulem}
\usepackage{amsfonts}

\usepackage{graphicx,color,hyperref}
\usepackage[caption=false]{subfig}
\hypersetup{colorlinks=true, linkcolor=blue, citecolor=blue, urlcolor=blue} 
\usepackage{svg}

\newcommand{\beq}[1]{\begin{equation}\label{#1}}
\newcommand{\eep}{\;.\end{equation}}
\newcommand{\eec}{\;,\end{equation}}
\newcommand{\eeq}{\end{equation}}

\newcommand{\ii}{\mathrm{i}}
\newcommand{\e}{\mathrm{e}}

\makeatletter
\newcommand{\fmarki}{*}
\newcommand{\fmarkii}{\ensuremath{\dagger}}
\newcommand{\fmarkiii}{\ensuremath{\ddagger}}
\newcommand{\fmarkiv}{\ensuremath{\mathsection}}
\newcommand{\fmarkv}{\ensuremath{\mathparagraph}}
\newcommand{\fmarkvi}{\ensuremath{\|}}
\newcommand{\fmarkvii}{**}
\newcommand{\fmarkviii}{\ensuremath{\dagger\dagger}}
\newcommand{\fmarkix}{\ensuremath{\ddagger\ddagger}}
                
\def\@fnsymbol#1{{\ifcase#1\or \fmarki\or \fmarkii\or \fmarkiii\or \fmarkiv\or \fmarkv\or \fmarkvi\or \fmarkvii\or \fmarkviii\or \fmarkix \else\@ctrerr\fi}}
\makeatother

\renewcommand{\fmarki}{$*$}
\renewcommand{\fmarkii}{$*$}
\renewcommand{\fmarkiii}{$*$}
\renewcommand{\fmarkiv}{$*$}
\renewcommand{\fmarkv}{$*$}
\renewcommand{\fmarkix}{$*$}

\newcommand{\p}{\phi}

\newcommand{\D}{\Delta}

\DeclareMathAlphabet{\mathcal}{OMS}{cmsy}{m}{n} 

\usepackage{amsmath}
\usepackage{amssymb}
\usepackage{xcolor}
\usepackage{bbm}
\usepackage{physics}
\usepackage{float}
\usepackage{graphicx}
\usepackage{dcolumn} 
\usepackage{bm} 
\usepackage{siunitx}
\usepackage{enumitem}  

\makeatletter
\renewcommand*{\fnum@figure}{{\normalfont\bfseries \figurename~\thefigure}}
\makeatother

\allowdisplaybreaks

\definecolor{orange}{rgb}{1,0.5,0}

\graphicspath{{./img/}}

\DeclareMathAlphabet{\mathcal}{OMS}{cmsy}{m}{n} 

\makeatletter
\newcommand{\specificthanks}[1]{\@fnsymbol{#1}}
\makeatother

\begin{document}

\raggedbottom

\title{Zak Phase Quantization without Global Symmetry through Destructive Interference}

\author{Geonwoo Hwang}
\author{Jun-Won Rhim}
\email{jwrhim@ajou.ac.kr}
\affiliation{Department of Physics, Ajou University, Suwon 16499, Republic of Korea}


\begin{abstract}
The Zak phase provides a fundamental geometric characterization of 
one-dimensional band structures and, when quantized by symmetry, serves 
as a bulk indicator of polarization and boundary phenomena. 
Conventionally, its exact quantization is tied to a global symmetry of the 
Bloch Hamiltonian, such as inversion or reflection symmetry. 
Here, we demonstrate a distinct mechanism by which the Zak phase remains 
exactly quantized even when the corresponding protecting symmetry is 
explicitly broken at the level of the full Hamiltonian. 
The mechanism is based on local support symmetry and destructive 
interference: symmetry-breaking degrees of freedom are coupled to a 
topological parent system in such a way that the relevant Bloch states 
cannot leak into them. 
The preserved states therefore retain their parent wave functions and 
inherit their quantized Zak phase exactly, despite the absence of the 
protecting global symmetry. 
We establish this mechanism for general one-dimensional insulators and 
demonstrate it in an inversion-broken minimal model and a 
reflection-broken armchair nanoribbon, where a quantized $\pi$ Zak phase 
and associated boundary-localized states persist. 
We further show that perturbations violating the exact 
destructive-interference condition continuously dequantize the Zak phase, 
while a finite regime with a near-$\pi$ Zak phase and boundary response 
can survive. 
Our results establish wave-function support, rather than global symmetry 
alone, as a mechanism for protecting quantized quantum geometry in 
one-dimensional systems.
\end{abstract}

\maketitle

\section{Introduction}

The Zak phase, the Berry phase accumulated by a Bloch state across the
one-dimensional Brillouin zone, is one of the fundamental geometric
quantities in band theory~\cite{berry1984quantal,zak1989berry,xiao2010berry,
chiu2016classification,cooper2019topological,atala2013direct,cardano2017detection}.
Beyond its role as a geometric characterization of Bloch wave functions,
it is intimately connected to physically observable properties of
one-dimensional insulators~\cite{thouless1983quantization,niu1984quantised,
kingSmith1993theory,vanderbilt1993electric,resta1994macroscopic,
ortiz1994macroscopic,marzari1997maximally,resta1998quantum}.
Most notably, the Zak phase of the occupied bands determines the electronic
polarization modulo a polarization quantum~\cite{kingSmith1993theory,
vanderbilt1993electric,resta1994macroscopic,ortiz1994macroscopic,
marzari1997maximally,resta1998quantum,hughes2011inversion,fuchs2021orbital}
and, when quantized, provides a bulk diagnostic of topologically distinct
insulating phases~\cite{ssh1979solitons,rice1982elementary,ryu2002topological,
delplace2011zak,hughes2011inversion,grusdt2013topological,xiao2014surface,
xiao2015geometric,gao2015determination,wang2016measurement,
chaunsali2017demonstrating,zhu2018zak,grusdt2019topological,
jiao2021experimentally,fuchs2021orbital,reid2022dynamically,li2023direct}.
Its nontrivial value is also closely associated with boundary phenomena,
providing a simple manifestation of the bulk--boundary correspondence in
one dimension~\cite{ssh1979solitons,ryu2002topological,delplace2011zak,
grusdt2013topological,xiao2014surface,xiao2015geometric,wang2016measurement,
rhim2017bulk,cardano2017detection,chaunsali2017demonstrating,
esmann2018topological,li2018elastic,zhu2018zak,xue2019second,
whittaker2019effect,wang2021zak,jiang2021fourband}.
For these reasons, the Zak phase has become a paradigmatic example of how
the geometry of quantum states can acquire direct physical and topological
significance across electronic, atomic, photonic, acoustic, mechanical,
and other engineered one-dimensional systems~\cite{kitagawa2012observation,
atala2013direct,xiao2015geometric,cardano2017detection,
chaunsali2017demonstrating,esmann2018topological,li2018elastic,
xue2019second,whittaker2019effect,cooper2019topological,
jiao2021experimentally,wang2021zak,jiang2021fourband,
reid2022dynamically,li2023direct}.

A crucial ingredient behind this topological interpretation is 
quantization of this quantity~\cite{zak1989berry,ryu2002topological,
chiu2016classification}. 
In a generic one-dimensional crystal, the Zak phase is a continuous 
geometric quantity and is not restricted to discrete values
~\cite{zak1989berry,xiao2010berry,fuchs2021orbital}. 
Quantization arises when additional constraints are imposed on the Bloch 
Hamiltonian~\cite{zak1989berry,hughes2011inversion,
chiu2016classification}. 
Spatial symmetries such as inversion or reflection can pin the Zak phase 
to $0$ or $\pi$ modulo $2\pi$
~\cite{zak1989berry,hughes2011inversion,xiao2014surface,
xiao2015geometric,gao2015determination,wang2016measurement,
rhim2017bulk,fuchs2021orbital}, while related quantization mechanisms occur 
in the presence of chiral or other symmetry constraints
~\cite{ssh1979solitons,ryu2002topological,chiu2016classification,
jiao2021experimentally}. 
Once quantized, the Berry phase cannot change continuously as long as the 
protecting symmetry and the relevant bulk gap are maintained
~\cite{ssh1979solitons,ryu2002topological,hughes2011inversion,
chiu2016classification,fuchs2021orbital}. 
This symmetry-based viewpoint underlies much of the conventional 
classification of one-dimensional topological insulators
~\cite{ssh1979solitons,ryu2002topological,hughes2011inversion,
chiu2016classification}: a global 
symmetry of the Bloch Hamiltonian constrains the geometry of its 
eigenstates, thereby quantizing a bulk invariant and, under appropriate 
boundary conditions, diagnosing boundary-localized states
~\cite{ryu2002topological,delplace2011zak,grusdt2013topological,
xiao2014surface,xiao2015geometric,wang2016measurement,rhim2017bulk,
cardano2017detection,chaunsali2017demonstrating,zhu2018zak,
whittaker2019effect,jiao2021experimentally}.

This familiar picture naturally raises a more basic question: 
is a global protecting symmetry actually necessary for an exactly 
quantized Zak phase?
At first sight, the answer appears to be affirmative. 
If the symmetry that quantizes the Berry phase is explicitly broken, the 
Bloch eigenstates are generally free to explore a larger region of Hilbert 
space, and their Zak phase can continuously depart from its symmetry-pinned 
value without closing the bulk gap. 
This expectation, however, implicitly assumes that the symmetry-breaking 
degrees of freedom necessarily participate in the wave functions carrying 
the geometric phase. 
There exists another possibility: a Hamiltonian may be globally 
symmetry broken while selected eigenstates remain completely confined to 
a subspace on which the original symmetry structure is preserved.

In this work, we establish such a mechanism for one-dimensional band 
topology. 
Building on the insulating formulation of local support symmetry introduced 
in our previous work~\cite{RhimLSS}, we show that this principle provides 
a distinct mechanism for the exact quantization of the Zak phase in the 
absence of the corresponding global protecting symmetry.
Starting from a parent insulator with a symmetry-quantized Zak phase, we 
enlarge the system by introducing additional orbitals and couplings that 
explicitly break the protecting symmetry of the full Hamiltonian.
The couplings are arranged, however, so that destructive interference 
prevents the relevant Bloch states from leaking into the added 
symmetry-breaking sector. 
These states consequently retain exactly the same wave functions as their 
parent counterparts on their nonzero support and inherit their geometric 
phase unchanged. 
The Zak phase can therefore remain pinned to $\pi$ even though the 
corresponding inversion or reflection symmetry is absent from the full 
Bloch Hamiltonian. 
In this sense, quantization is protected not by a global symmetry of the 
entire system, but by a local support symmetry acting on the portion 
of Hilbert space actually occupied by the topological states.

We first formulate this principle for a general one-dimensional insulator 
and identify the destructive-interference condition that guarantees exact 
inheritance of the parent Bloch eigenstates and their Zak phases. 
We then construct a minimal two-orbital parent model with an 
inversion-quantized $\pi$ Zak phase and show that coupling it to an 
inversion-breaking auxiliary orbital can preserve the topological band 
exactly. 
Importantly, the construction does not require the complementary parent 
band to be flat: finite-range couplings realizing the required destructive 
interference can also be constructed when the unoccupied band is 
dispersive. 
We further demonstrate the mechanism in a six-site armchair-nanoribbon 
model, where a reflection-protected occupied topology survives after the 
reflection symmetry of the enlarged lattice is explicitly broken. 
In both settings, the exactly quantized Zak phase is accompanied by 
boundary-localized states in finite systems. 
Finally, by introducing perturbations that open leakage channels out of 
the local-support sector, we show how exact quantization evolves into a 
finite regime of near-quantized geometric response.

Our results thus provide a route to Zak-phase quantization that lies 
outside the conventional paradigm of global symmetry protection. 
They show that the relevant question for geometric quantization need not 
be whether the entire Hamiltonian possesses a protecting symmetry, 
but whether the wave-function support carrying the geometric phase 
can remain isolated from symmetry-breaking degrees of freedom. 
This support-based perspective enlarges the design space of 
one-dimensional topological insulators and suggests destructive 
interference as a general tool for retaining quantized quantum geometry 
in globally symmetry-broken systems.



\section{Local-support protection of one-dimensional band topology}
\label{sec:lss_zak_general}

We begin by recasting the insulating construction of Ref.~\cite{RhimLSS} in a form adapted to one-dimensional band topology.  The central question is the following.  Suppose that a set of bands of a parent one-dimensional insulator carries a quantized Zak phase protected by a symmetry.  Can the corresponding topology survive after the parent system is enlarged by additional orbitals in such a way that the protecting symmetry is no longer a symmetry of the full Bloch Hamiltonian?  The local-support construction provides an affirmative answer when the coupling to the added degrees of freedom annihilates the topological Bloch states.  In this case, the relevant eigenstates remain entirely supported on the parent sector due to destructive interference, and hence retain both their wave functions and their geometric phases even though the enlarged Hamiltonian need not possess the original global symmetry.

Consider a one-dimensional Bloch Hamiltonian decomposed into a parent sector $A$ and an auxiliary symmetry-breaking sector $B$,
\begin{equation}
H(k)=
\begin{pmatrix}
 h_A(k) & W(k)\\
 W^{\dagger}(k) & h_B(k)
\end{pmatrix},
\label{eq:block_H_1d}
\end{equation}
where $h_A(k)$ is an $N_A\times N_A$ Hamiltonian and $h_B(k)$ describes $N_B$ additional orbitals.  
We assume that $h_A(k)$ is an insulator containing one or more bands whose
Zak phases are quantized by a symmetry of the parent Hamiltonian.
Let $|u_n(k)\rangle$ denote a parent Bloch eigenstate to be protected,
satisfying
\begin{equation}
h_A(k)|u_n(k)\rangle
=
\varepsilon_n(k)|u_n(k)\rangle.
\end{equation}
The key LSS condition is
\begin{equation}
 W^{\dagger}(k)|u_n(k)\rangle=0
 \label{eq:annihilation_condition}
\end{equation}
for every $k$ and for each band $n$ to be protected.
Whenever Eq.~\eqref{eq:annihilation_condition} holds, the enlarged Hamiltonian possesses the exact eigenstates
\begin{equation}
 |U_n(k)\rangle=
 \begin{pmatrix}
 |u_n(k)\rangle\\[2pt]
 0
 \end{pmatrix},
 \qquad
 H(k)|U_n(k)\rangle
 =\varepsilon_n(k)|U_n(k)\rangle.
\label{eq:embedded_state}
\end{equation}
Thus, turning on $W(k)$ does not perturb these bands at any order: neither their energies nor their Bloch eigenvectors are modified.  The states in Eq.~\eqref{eq:embedded_state} have exactly vanishing amplitude on the added orbitals denoted by $B$.  
In real space, Eq.~\eqref{eq:annihilation_condition} is naturally interpreted as a destructive-interference constraint: all hopping amplitudes from the protected state into the auxiliary sector cancel.
%

This observation has a particularly transparent consequence in one dimension.  For an isolated occupied band, the Zak phase is
\begin{equation}
 \gamma
 =i\int_{\mathrm{BZ}} dk\,
 \langle u(k)|\partial_k u(k)\rangle
 \quad (\mathrm{mod}\ 2\pi).
\label{eq:zak_parent}
\end{equation}
Because the embedding in Eq.~\eqref{eq:embedded_state} is $k$ independent,
\begin{equation}
 i\langle U_n(k)|\partial_k U_n(k)\rangle
 =i\langle u_n(k)|\partial_k u_n(k)\rangle,
\label{eq:berry_connection_inheritance}
\end{equation}
and therefore
\begin{equation}
 \gamma_{H,n}=\gamma_{A,n}
 \quad (\mathrm{mod}\ 2\pi),
\label{eq:zak_inheritance}
\end{equation}
where $\gamma_{H,n}$ and $\gamma_{A,n}$ are the Zak phases corresponding to $|U_n(k)\rangle$ and $|u_n(k)\rangle$, respectively.
The same statement applies to a group of occupied bands: if the complete occupied subspace relevant to the insulating gap is embedded according to Eq.~\eqref{eq:embedded_state}, its non-Abelian Berry connection, Wilson loop, and total occupied Zak phase are inherited unchanged from the parent sector.  Consequently, a parent value such as $\gamma_A=\pi$, the total Zak phase of the occupied bands, remains exactly pinned to $\pi$ as long as the protected occupied subspace remains spectrally isolated and Eq.~\eqref{eq:annihilation_condition} is maintained.

Importantly, the symmetry responsible for quantizing $\gamma_A$ need not survive as a global symmetry of $H(k)$.  Suppose, for example, that $h_A(k)$ obeys a crystalline symmetry $g$ that quantizes its Zak phase, while $h_B(k)$ and/or $W(k)$ explicitly violate the corresponding symmetry of the enlarged system. Equation~\eqref{eq:annihilation_condition} nevertheless prevents the symmetry-breaking sector from entering the protected Bloch states.  The symmetry then acts effectively only on the support of those states.  In this sense, the quantization inherited from the parent problem is protected by a local support symmetry: the full Hamiltonian can be globally symmetry broken, while the wave-function sector carrying the topological phase remains confined to the part of the Hilbert space on which the original symmetry is well defined.  The mechanism should therefore be distinguished from an emergent global symmetry of the enlarged Hamiltonian; what survives exactly is the symmetry-constrained eigenstate structure of the parent sector.

The coupling condition can be expressed directly in terms of the complementary subspace of $h_A(k)$.  Introduce the occupied projector
\begin{equation}
 P_A(k)=\sum_{n\in\mathrm{occ}}
 |u_n(k)\rangle\langle u_n(k)|,
\end{equation}
and $Q_A(k)=1-P_A(k)$.  Equation~\eqref{eq:annihilation_condition} is equivalent to
\begin{equation}
 W^{\dagger}(k)P_A(k)=0,
\label{eq:projector_condition}
\end{equation}
so every row of $W^{\dagger}(k)$ must lie in the complementary subspace selected by $Q_A(k)$.  If $\{|v_a(k)\rangle\}$ spans the unoccupied subspace of $h_A(k)$, one may write
\begin{equation}
 W^{\dagger}(k)
 =\sum_a |\chi_a(k)\rangle\langle v_a(k)|,
\label{eq:W_unoccupied}
\end{equation}
with coefficients collected into vectors $|\chi_a(k)\rangle$ in the auxiliary sector.  This form makes the orthogonality mechanism explicit: the auxiliary orbitals couple only to directions in the parent Hilbert space that are orthogonal to the protected occupied subspace.

For a short-ranged tight-binding model, however, Eq.~\eqref{eq:W_unoccupied} is useful only when the resulting matrix elements can be represented by finite Laurent polynomials in $z=e^{ik}$.  A convenient route is therefore to choose the complementary eigenvector entering Eq.~\eqref{eq:W_unoccupied} so that an unnormalized representative has finite support in Bloch-phase space.  Flat bands provide a systematic realization of this property: for a finite-range Bloch Hamiltonian, an unnormalized flat-band eigenvector can be chosen, under the usual nonsingular conditions, as a finite combination of Bloch factors~\cite{rhim2019classification,rhim2021singular,kim2023general}.  Its components can then be used directly to construct finite-range hopping amplitudes satisfying Eq.~\eqref{eq:annihilation_condition}.  In real space, the same algebra appears as exact cancellation of hopping paths into the newly attached sites.

The one-dimensional setting also exposes a minimal version of the construction.  For a two-band parent Hamiltonian with one occupied state $|u(k)\rangle$ and one unoccupied state $|v(k)\rangle$, orthogonality fixes the allowed coupling direction uniquely up to an overall function.  For a single added orbital, one may take
\begin{equation}
 W^{\dagger}(k)=\lambda(k)\langle v(k)|,
\label{eq:two_band_coupling}
\end{equation}
where $\lambda(k)$ is chosen so that $W(k)$ corresponds to the desired finite-range hopping structure.  Then
\begin{equation}
 W^{\dagger}(k)|u(k)\rangle
 =\lambda(k)\langle v(k)|u(k)\rangle=0
\end{equation}
identically.  This provides a direct design principle for the models studied below: start from a one-dimensional parent insulator with a quantized Zak phase, identify the orthogonal Bloch direction, and attach symmetry-breaking orbitals only through that direction.  The resulting Hamiltonian can violate the parent crystalline symmetry globally while retaining the parent topological eigenstate exactly.

There are two logically distinct requirements in this construction.  First, the wave-function condition, Eq.~\eqref{eq:annihilation_condition}, guarantees exact inheritance of the parent Bloch states and hence of their Zak phase.  Second, the spectral condition requires that these states continue to define the occupied band or occupied band subspace associated with the insulating gap of interest.  The latter can fail if auxiliary bands cross the Fermi level or invert with the protected bands, even while Eq.~\eqref{eq:annihilation_condition} remains exact.  Separating these two requirements is useful in what follows: the destructive-interference condition controls the eigenvectors, whereas the placement of the auxiliary spectrum controls whether the inherited geometric phase is the physically relevant occupied-band invariant.

This reformulation specializes the general insulating local-support mechanism of Ref.~\cite{RhimLSS} to one-dimensional Berry-phase topology.  It also provides the starting point for our subsequent analysis, where we ask how far the exact condition can be relaxed and which symmetry-breaking coupling patterns preserve, or approximately preserve, the parent value of the Zak phase.

\begin{figure*}[!t]
    \centering
    \includegraphics[width=0.96\textwidth]{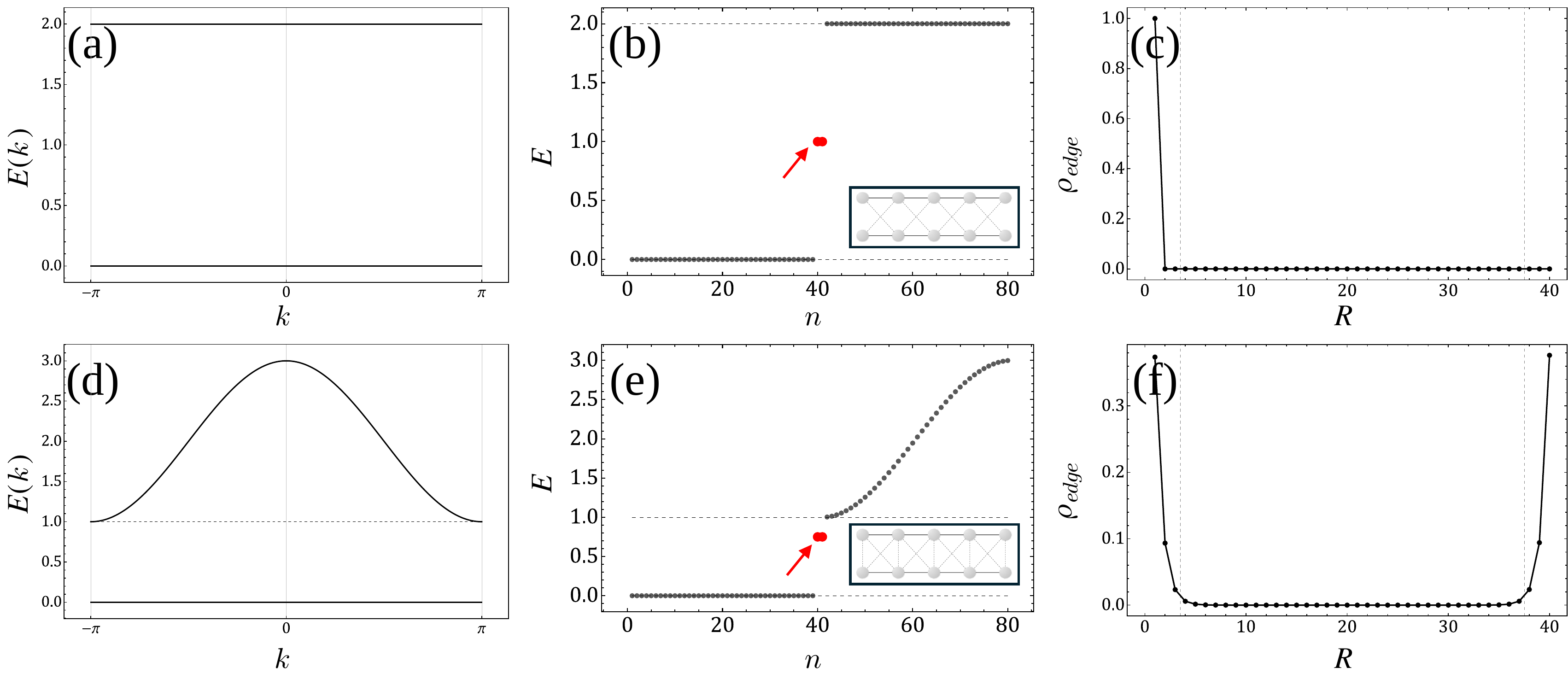}
   \caption{
    Bulk and boundary properties of the two $2\times2$ parent models.
Panels (a)--(c) correspond to the Perfect Pythagorean model with
$(a_0,b_0,\theta,\eta)=(1,0,\pi/4,1)$, while panels (d)--(f)
correspond to the Imperfect Pythagorean model with
$(a_0,b_0,\theta,\eta)=(1,0.5,\pi/4,1)$.
Panels (a,d) show the bulk band structures under periodic boundary
conditions, and panels (b,e) show the finite-chain spectra under open
boundary conditions, with the in-gap edge modes highlighted in red. The lattice and hopping structures are illustrated in the insets.
Panels (c,f) display the real-space probability densities of the
corresponding edge modes, demonstrating their localization at the
boundaries. In both models, the occupied band carries the quantized
Zak phase $\gamma_{\rm occ}=\pi$.}
    \label{fig:fig1}
\end{figure*}

\section{Two-orbital insulators with LSS}
\label{sec:one_dimensional_lss}

We construct a one-dimensional insulator in which the Zak phase of an individual band remains exactly quantized even though the crystalline symmetry responsible for its quantization is absent from the full Hamiltonian.  The construction begins with a two-band inversion-symmetric parent model having an occupied flat band with Zak phase $\pi$.  We then couple the parent to additional orbitals that break inversion globally.  When the coupling annihilates the occupied parent eigenvector, destructive interference prevents the state from leaking into the symmetry-breaking sector.  The resulting eigenstate is identical to the parent state on its support and therefore inherits its quantized Zak phase. This realizes the insulating version of local support symmetry (LSS).

\subsection{Inversion-symmetric parent model with a flat band}
\label{subsec:parent_family}

Consider a one-dimensional two-orbital tight-binding model with hoppings up to neighboring unit cells,
\begin{equation}
 h_1(k)=d_0(k)\sigma_0+\bm d(k)\cdot\bm\sigma,
 \label{eq:parent_hamiltonian_section}
\end{equation}
where $\bm d=(d_x,d_y,d_z)$ is a real-valued vector function, $\sigma_0$ is an identity matrix, and $\sigma_\alpha$, with $\alpha \in \{x,y,z\}$, is a Pauli matrix.
We consider a representative two-orbital parent model in which the two orbitals are exchanged by inversion. Choosing the unit cell and orbital phases such that the inversion center lies midway between them, the inversion operator is represented by $P=\sigma_x$.  The symmetry condition
\begin{equation}
 P h_1(k)P^{-1}=h_1(-k)
 \label{eq:inversion_condition_section}
\end{equation}
requires $d_0$ and $d_x$ to be even in $k$, whereas $d_y$ and $d_z$ are odd.  The most general nearest-neighbor form consistent with these parities is
\begin{align}
 d_0(k)&=a_0+a_1\cos k,
 &d_x(k)&=b_0+b_1\cos k,\nonumber\\
 d_y(k)&=c_1\sin k,
 &d_z(k)&=d_1\sin k.
 \label{eq:nearest_neighbor_parent_section}
\end{align}
The corresponding band energies are given by $E_\pm(k)=d_0(k)\pm |\bm d(k)|$.

For the LSS analysis, it is beneficial if one of the bands is flat.
We choose the energy origin such that the lower band is flat at $E_-(k)=0$.  Assuming $d_0(k)>0$, this condition is equivalent to
\begin{equation}
 d_0^2(k)-|\bm d(k)|^2=\det h_1(k)=0
 \label{eq:flatness_determinant_section}
\end{equation}
for every momentum.  
Introducing $C=c_1^2+d_1^2$, Eq.~\eqref{eq:flatness_determinant_section} leads to three conditions (i) $a_0^2-b_0^2=C$, (ii) $a_0a_1=b_0b_1$, and (iii) $b_1^2-a_1^2=C$. 
Then, our gapped topological family is obtained by taking
$a_1=\eta b_0$, $b_1=\eta a_0$, $\eta=\pm1$, and $a_0^2-b_0^2=c_1^2+d_1^2$ with $a_0>|b_0|$.
Thus
\begin{align}
 d_0(k)&=a_0+\eta b_0\cos k,
 &d_x(k)&=b_0+\eta a_0\cos k,\nonumber\\
 d_y(k)&=c_1\sin k,
 &d_z(k)&=d_1\sin k.
 \label{eq:one_flat_band_family_section}
\end{align}
In this case, the dispersion of the upper band is given by $E_+(k)=2\left(a_0+\eta b_0\cos k\right)$.
The minimum direct separation between the two bands is $\Delta_{\mathrm{dir}}=2(a_0-|b_0|)>0$.

We refer to the special case of Eq.~\eqref{eq:one_flat_band_family_section} with $b_0=0$ as the perfect Pythagorean case.
In this case, the coefficients satisfy a relation $a_0^2=c_1^2+d_1^2$ and both bands are flat with energies $E_-(k)=0$ and $E_+(k)=2a_0$.  It is therefore a doubly-flat limit of the general one-flat-band family.

\subsection{Quantized Zak phase of the parent band}
\label{subsec:parent_zak}

For an isolated normalized eigenstate $|u_-(k)\rangle$, the Zak phase is
\begin{equation}
 \gamma_-=\int_{-\pi}^{\pi}dk\,
 \ii\langle u_-(k)|\partial_k u_-(k)\rangle
 \quad (\mathrm{mod}\ 2\pi).
 \label{eq:zak_definition_section}
\end{equation}
In an inversion-symmetric one-dimensional system, it satisfies
$\e^{\ii\gamma_-}=\xi_-(0)\xi_-(\pi)$, where $\xi_-(k_\star)=\pm1$ is the inversion eigenvalue of the lower band at $k_\star=0,\pi$.  At these momenta the odd components $d_y$ and $d_z$ vanish, and $h_1(k_\star)=d_0(k_\star)\sigma_0+d_x(k_\star)\sigma_x$.
The inversion eigenvalue of the lower band is therefore $\xi_-(k_\star)=-\operatorname{sgn}d_x(k_\star)$.  
For Eq.~\eqref{eq:one_flat_band_family_section}, $d_x(0)=b_0+\eta a_0$ and $d_x(\pi)=b_0-\eta a_0$, and hence
$d_x(0)d_x(\pi)=b_0^2-a_0^2<0$.
The inversion characters at $0$ and $\pi$ are opposite, yielding
\begin{equation}
 \gamma_-=\pi\quad(\mathrm{mod}\ 2\pi).
 \label{eq:parent_pi_zak_section}
\end{equation}


The quantized Zak phase also admits a geometric interpretation closely related to the parameter-space loop introduced by Ryu and Hatsugai~\cite{ryu2002topological}.  
For the present family, the pseudospin vector is $\bm d(k)=\left(b_0+\eta a_0\cos k,\,c_1\sin k,\,d_1\sin k\right)$.
Although all three components can be nonzero, the $y$ and $z$ components are proportional to the same function $\sin k$.  Consequently, $\bm d(k)$ traces an ellipse in the plane spanned by $\hat{\bm e}_x=(1,0,0)$ and $\hat{\bm e}_{\perp}=
(0,c_1,d_1)/\sqrt{c_1^2+d_1^2}$, which is a plane passing through the origin of the pseudospin space.  
The ellipse is displaced from the origin by $b_0\hat{\bm e}_x$ and has semiaxis $a_0$ along $\hat{\bm e}_x$.  
Hence, the condition $a_0>|b_0|$ implies that the ellipse encloses the origin once.  
Equivalently, the normalized vector $\hat{\bm d}(k)$ winds once along the corresponding great circle of the Bloch sphere, giving a Berry phase of $\pi$. The Zak phase quantization proof in this manner can be found in appendix \ref{app:direct_phase_cal}, which separates Zak phase into two terms: a winding contribution and an additional $d_z(k)$-dependent contribution.

This planar-loop structure can also be expressed through the operator $\Gamma=\hat{\bm n}\cdot\bm\sigma$, where $\hat{\bm n}
=(0,-d_1,c_1)/\sqrt{c_1^2+d_1^2}$ is a normal vector to the plane of the loop.  
The traceless part of the parent Hamiltonian, $\widetilde h_1(k)=h_1(k)-d_0(k)\sigma_0=\bm d(k)\cdot\bm\sigma$, then satisfies $\{\Gamma,\widetilde h_1(k)\}=0$.
The full parent Hamiltonian does not, in general, possess this chiral symmetry because of the momentum-dependent scalar term
$d_0(k)\sigma_0$.  
Nevertheless, this term does not affect the Bloch eigenvectors, so the planar-loop geometry and the resulting Zak phase remain unchanged.  
In the present inversion-symmetric parent model, the same nontrivial geometry is equivalently diagnosed by $d_x(0)d_x(\pi)<0$, or $\xi_-(0)\xi_-(\pi)=-1$.  
Thus, the Ryu--Hatsugai loop picture provides the geometric interpretation of the parent-band topology, whereas the LSS construction developed below preserves the corresponding eigenvector and its $\pi$ Zak phase even after inversion is broken in the enlarged Hamiltonian.

Fig.~\ref{fig:fig1} compares the bulk and boundary spectra of representative Perfect and Imperfect Pythagorean parent models. For the perfect Pythagorean case, we choose $(a_0,b_0,\theta,\eta)=(1,0,\pi/4,1)$, which yields two perfect flat bands at $E_-=0$ and $E_+=2a_0$, as shown by the periodic-boundary spectrum in Fig.~\ref{fig:fig1}(a). By contrast, the Imperfect case in Fig.~\ref{fig:fig1}(d), evaluated for $(a_0,b_0,\theta,\eta)=(1,0.5,\pi/4,1)$, retains the flat occupied band at $E_-=0$ while the upper band disperses according to $E_+(k)=2(a_0+\eta b_0\cos k)$. The corresponding spectra of finite open chains are displayed in Figs.~\ref{fig:fig1}(b) and \ref{fig:fig1}(e), where the gray and red points represent the bulk and edge-localized states, respectively. The real-space lattice and hopping structures are plotted in the insets, where one can identify the inversion symmetry. To verify the boundary character of the selected states, Figs.~\ref{fig:fig1}(c) and \ref{fig:fig1}(f) show their cell-resolved densities, $\rho_{\mathrm{edge}}(R)=\sum_{j=1}^{2}|\psi_{R,j}|^2$, which are concentrated near one or both ends of the chain and decay into the bulk. The appearance of these in-gap boundary-localized states is consistent with the quantized occupied-band Zak phase $\gamma_{\mathrm{occ}}=\pi$ in both parent models.

\begin{figure*}[!t]
    \centering
    \includegraphics[width=0.96\textwidth]{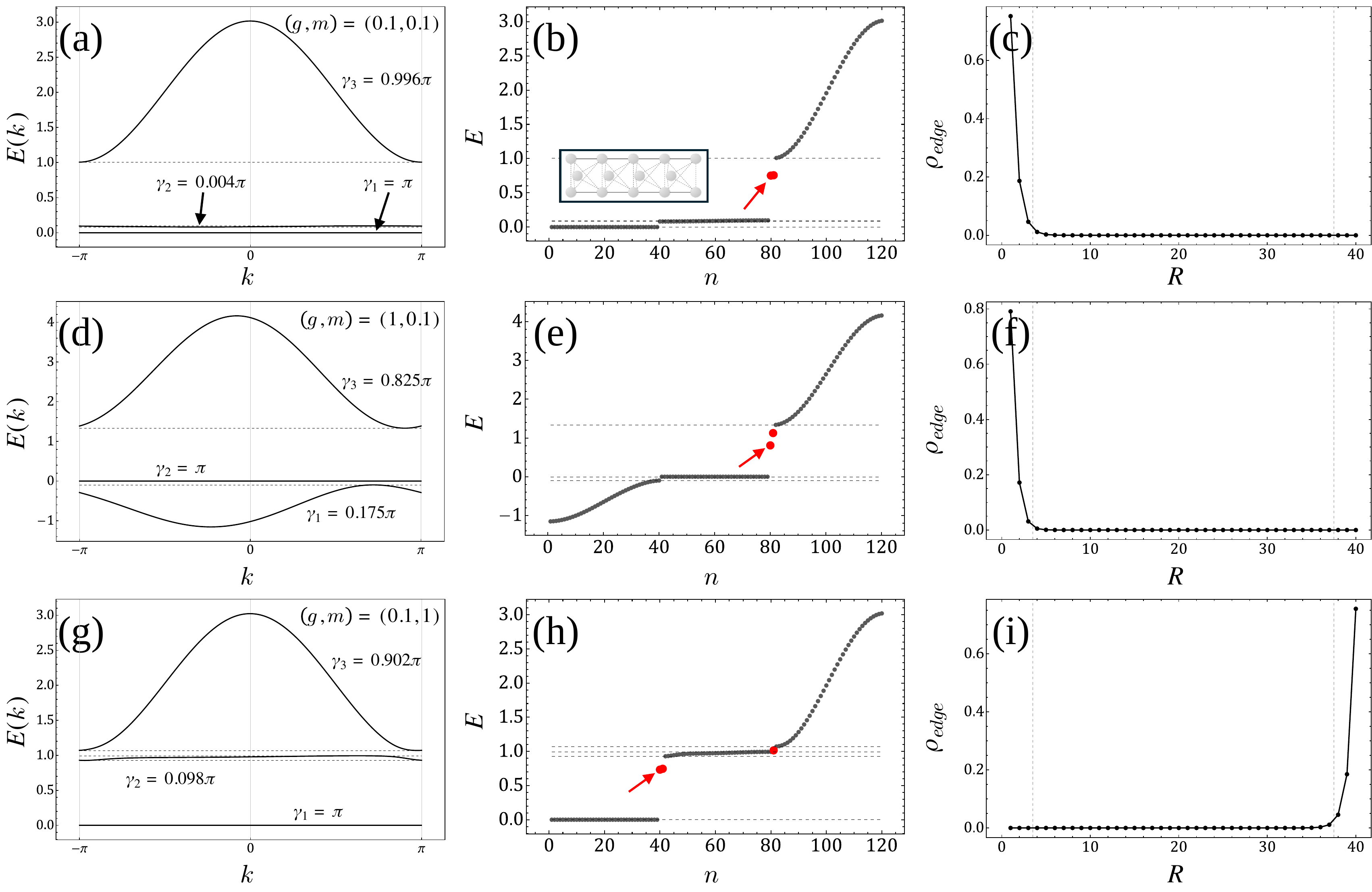}
    \caption{
    Band structure, OBC energy, and edge mode density plot for symmetry broken model. Rows (a)--(c), (d)--(f), and (g)--(i) correspond to the three representative LSS parameter sets $(g,m)=(0.1,0.1)$, $(1,0.1)$, $(0.1,1)$, respectively. Panels (a), (d), and (g) show the three PBC bands, ordered by increasing energy; the horizontal dashed lines mark the adjacent bulk gaps.
    Panels (b), (e), and (h) show the corresponding spectra of finite open chains. Gray points denote all OBC eigenvalues, whereas red points denote the selected boundary-localized states within the indicated bulk gap. The inset in panel (b) specifies the real-space unit cell and hopping convention used for all calculations. Panels (c), (f), and (i) display the cell-resolved densities of the selected states, $\rho_{\mathrm{edge}}(R)=\sum_{j=1}^{3}|\psi_{R,j}|^2$, demonstrating localization near one or both boundaries and decay into the bulk.}
    \label{fig:fig2}
\end{figure*}

\subsection{Exact LSS embedding}
\label{subsec:lss_embedding}

We next couple the parent sector $S_1$ to a sector $S_2$ that need not preserve inversion.  The enlarged Hamiltonian is
\begin{equation}
 H(k)=
 \begin{pmatrix}
 h_1(k)&h_{12}(k)\\
 h_{21}(k)&h_2(k)
 \end{pmatrix},
 \qquad
 h_{12}(k)=h_{21}^\dagger(k).
 \label{eq:lss_block_hamiltonian_section}
\end{equation}
Following the general insulating theory of LSS, an occupied parent eigenstate $v^{(o)}(k)$ remains an exact eigenstate of the full Hamiltonian if
\begin{equation}
 h_{21}(k)v^{(o)}(k)=0
 \label{eq:lss_kernel_section}
\end{equation}
for every $k$.  Under this condition,
\begin{equation}
 V^{(o)}(k)=
 \begin{pmatrix}v^{(o)}(k)\\0\end{pmatrix}
 \label{eq:embedded_state_section}
\end{equation}
satisfies $H(k)V^{(o)}(k)=0$.  The state has strictly zero amplitude on $S_2$ and is therefore compactly supported on the parent sector.  Since its nonzero components are identical to those of the parent eigenstate, the embedded state retains the exact Zak phase $\gamma=\pi$ even though inversion is not a symmetry of $H(k)$.

For the present two-band parent, this construction can be made explicit.  Write
\begin{equation}
 h_1(k)=
 \begin{pmatrix}
 p_+(k)&q(k)\\
 q^*(k)&p_-(k)
 \end{pmatrix},
\end{equation}
where $q=d_x-\ii d_y$ and $p_\pm=d_0\pm d_z$.
The flat-band condition in \eqref{eq:flatness_determinant_section} gives $|q(k)|^2=p_+(k)p_-(k)$.
Convenient unnormalized occupied and unoccupied eigenvectors are
\begin{equation}
 v^{(o)}(k)=
 \begin{pmatrix}q(k)\\-p_+(k)\end{pmatrix},
 \qquad
 v^{(u)}(k)=
 \begin{pmatrix}q(k)\\p_-(k)\end{pmatrix}.
 \label{eq:parent_vectors_section}
\end{equation}
Then, one directly verifies $h_1v^{(o)}=0$, $h_1v^{(u)}=2d_0v^{(u)}$, and $v^{(u)\dagger}v^{(o)}=|q|^2-p_+p_-=0$.
Because the orthogonal complement of $v^{(o)}$ is one dimensional, every row of an allowed coupling matrix can be chosen as
\begin{equation}
 [h_{21}(k)]_{s,:}=\lambda_s(k)v^{(u)\dagger}(k),
 \label{eq:h21_construction_section}
\end{equation}
where $s$ labels orbitals in $S_2$.  Equation~\eqref{eq:lss_kernel_section} then follows immediately from the above results.

For generic $b_0\neq0$, $v^{(u)}$ is the eigenvector of the dispersive upper band rather than an unoccupied flat band.  Nevertheless, the unnormalized components
\begin{align}
 q(k)&=b_0+\eta a_0\cos k-\ii c_1\sin k,\nonumber\\
 p_-(k)&=a_0+\eta b_0\cos k-d_1\sin k
 \label{eq:explicit_unoccupied_vector_section}
\end{align}
are finite Laurent polynomials in $\e^{\ii k}$.  Therefore Eq.~\eqref{eq:h21_construction_section} can be implemented using finite-range real-space hoppings.  In the perfect Pythagorean case ($b_0=0$), the upper band also becomes flat, and the construction reduces to the unoccupied-flat-band prescription of the general LSS theory.  The present two-band example thus shows that an unoccupied flat band is a sufficient, but not necessary, route to a local LSS coupling.

\begin{figure*}[!t]
    \centering

    \includegraphics[width=0.96\textwidth]{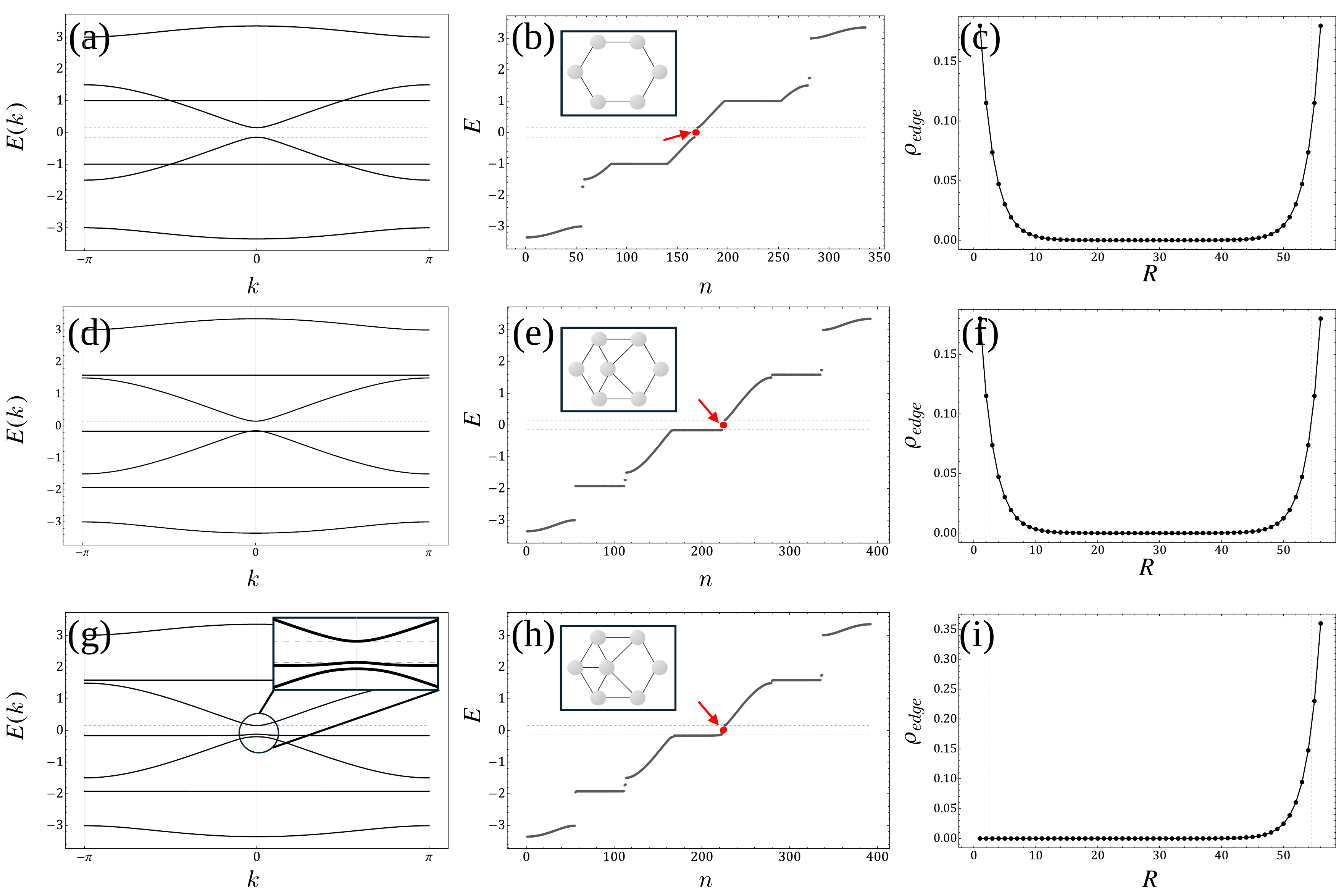}

    \caption{
    Band structures, finite-chain spectra, and boundary-state densities for
the parent armchair nanoribbon and its LSS-derived models.
Panels (a)--(c), (d)--(f), and (g)--(i) correspond to the reflection-symmetric
parent model, the ideal LSS model, and the perturbed LSS model, respectively.
Panels (a), (d), and (g) show the PBC band structures over the Brillouin zone
$k\in[-\pi,\pi]$. The inset in panel (g) highlights the weak dispersion of
the LSS-derived band induced by the leakage perturbation.
Panels (b), (e), and (h) show the corresponding finite-chain spectra under
open boundary conditions. Gray points denote all OBC eigenvalues, while red
points indicate the selected in-gap boundary-localized states. The lattice
insets illustrate the real-space unit cells and hopping configurations of
the corresponding models.
Panels (c), (f), and (i) show the cell-resolved probability densities of the
selected in-gap states, demonstrating their localization near the boundaries
and decay into the bulk.}
    \label{fig:armchair_spectra}
\end{figure*}

Fig.~\ref{fig:fig2} demonstrates how the topological parent band evolves after coupling the Imperfect Pythagorean model to an additional inversion-breaking orbital. All three examples use the same parent parameters, $a_0=1$, $b_0=0.5$, $\eta=1$, and $\theta=\pi/4$, with $\alpha=\sqrt{a_0^2-b_0^2}$, $c_1=\alpha\cos\theta$, and $d_1=\alpha\sin\theta$, while the coupling strength $g$ and the onsite energy $m$ of the added orbital are varied, such that Fig.~\ref{fig:fig2}(a)--(c),
\ref{fig:fig2}(d)--(f), and
\ref{fig:fig2}(g)--(i) correspond to
$(g,m)=(0.1,0.1)$, $(1,0.1)$, and $(0.1,1)$, respectively.
For $(g,m)=(0.1,0.1)$, the weakly coupled additional orbital leaves the LSS-protected band as the lowest band, with the three Zak phases given by $(\gamma_1,\gamma_2,\gamma_3)/\pi=(1.000,0.004,0.996)$, from the lowest band. Increasing the coupling to $(g,m)=(1,0.1)$ rearranges the bands so that the exactly quantized LSS band becomes the middle band, yielding $(\gamma_1,\gamma_2,\gamma_3)/\pi=(0.175,1.000,0.825)$. For $(g,m)=(0.1,1)$, the protected band again forms the lowest band, with $(\gamma_1,\gamma_2,\gamma_3)/\pi=(1.000,0.098,0.902)$. Thus, although inversion is broken and the Zak phases of the other bands are generally nonquantized, the band whose eigenvector has the form $V^{(o)}(k)=(v^{(o)}(k),0)^{T}$ retains its exact $\pi$ Zak phase as a consequence of LSS. 

\subsection{Destructive-interference mechanism}
\label{subsec:destructive_interference_section}

The destructive-interference mechanism is already manifest in the zero component of the extended eigenvector in Eq.~\eqref{eq:embedded_state_section}.  Indeed, the eigenvalue equation in the symmetry-breaking sector reads $h_{21}(k)v^{(o)}(k)+h_2(k)0=0$.
Although the individual hoppings contained in $h_{21}(k)$ are generally nonzero, their amplitudes cancel when acting on $v^{(o)}(k)$, so that the resulting state has exactly zero weight on $S_2$.  This mechanism differs from the familiar compact localized states of flat-band systems: $V^{(o)}(k)$ is an extended Bloch eigenstate, rather than a state confined to a finite real-space cluster, and its locality refers instead to its support in the orbital subspace $S_1$.  The parent inversion operation therefore remains well defined on the nonzero support of the state even though inversion is absent from the full Hamiltonian.

The exact quantization relies on this destructive-interference condition.
A perturbation $\delta h_{21}$ satisfying $\delta h_{21}(k)v^{(o)}(k)\neq0$ allows the state to acquire a finite component in $S_2$ and thereby modifies its Bloch eigenvector. Since inversion is not a symmetry of the
full Hamiltonian, no global symmetry then enforces quantization of the
Zak phase, which generally departs from $\pi$. Thus, in the enlarged
system, the exact $\pi$ Zak phase is tied directly to the destructive
interference that produces the zero component of $V^{(o)}(k)$.

\begin{figure*}[!t]
    \centering
    \includegraphics[width=0.96\textwidth]{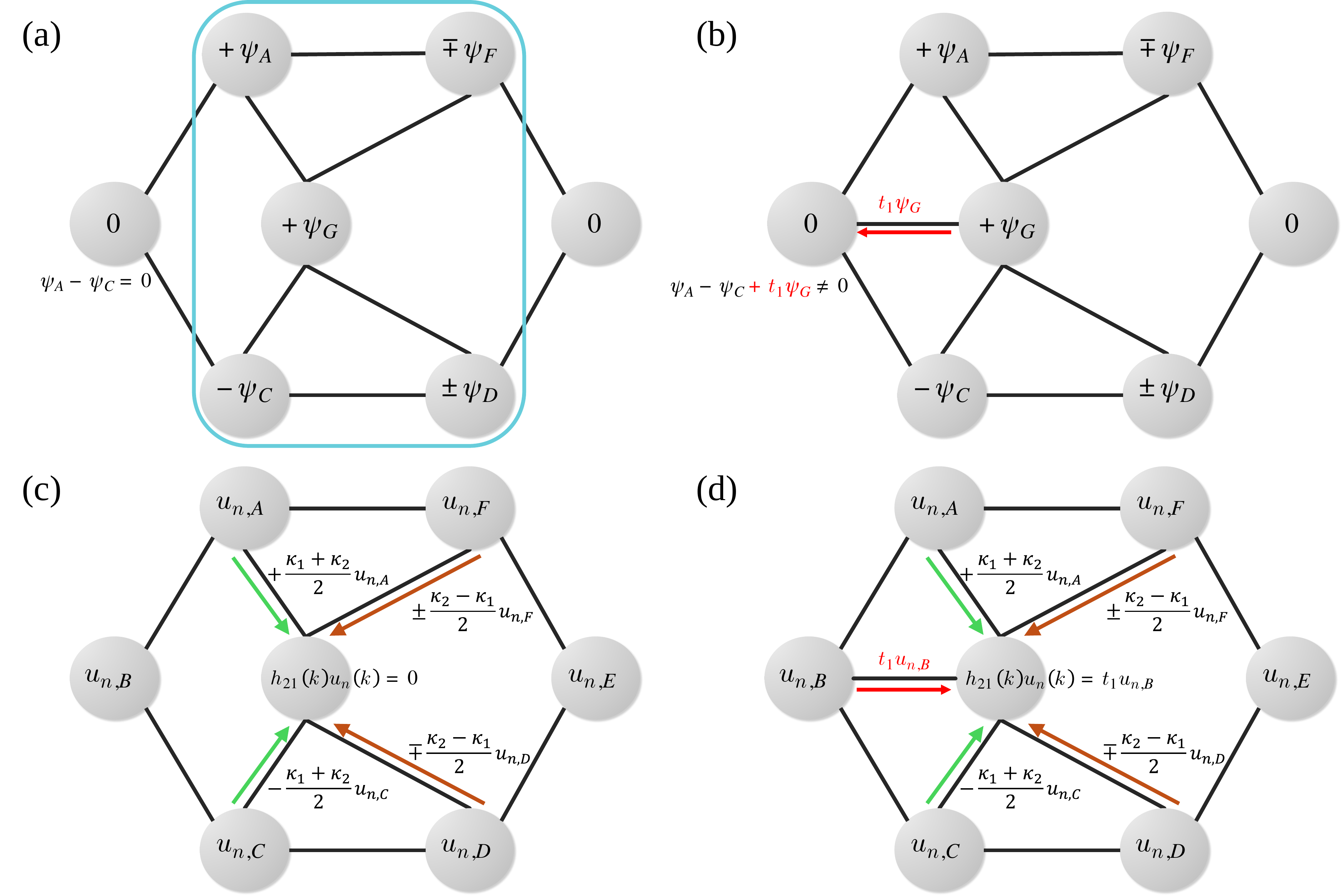}
    \caption{Real-space illustration of CLS and the destructive-interference mechanism and their
    breakdown under a leakage perturbation for $t_1\neq0$ and $t_2=0$.
    (a) In the ideal LSS construction, hopping amplitudes from the parent sector to connector sites $B$ and $E$ cancel by destructive interference, producing nodes at these sites and allowing the flat-sector eigenstates, including their amplitudes on the auxiliary site $G$, to remain compactly localized within a single unit cell. 
    (b) The additional $G$--$B$ hopping $t_1$  introduces an uncanceled amplitude $t_1\psi_G$ from $G$ to $B$, coupling the original flat-band sector to the remaining states.
    (c) In the ideal LSS construction, hopping amplitudes from the parent sector
    to the auxiliary symmetry-breaking site $G$ cancel exactly by destructive interference, leaving
    the LSS-protected state with zero amplitude on $G$.
    (d) The additional $G$--$B$ hopping $t_1$ opens a leakage channel and violates
    the kernel condition in Eq.~\eqref{eq:armchair_lss_kernel}.
    Consequently, the parent-derived state is no longer an exact eigenstate of the perturbed Hamiltonian, allowing its Zak phase to depart from the quantized value and the nominally flat band to acquire a finite dispersion.}
    \label{fig:fig4}
\end{figure*}

\section{Six-site armchair nanoribbon with local support symmetry}
\label{sec:six_site_armchair}

We now demonstrate the LSS construction in a narrow armchair nanoribbon containing six sites per unit cell.  This model provides a lattice realization in which a reflection-protected parent topology survives after the reflection symmetry is broken globally.  We first introduce a simple reflection-symmetric parent with two exact flat bands, and then add an auxiliary orbital whose couplings break the global reflection while preserving selected parent eigenvectors through destructive interference.  Finally, we examine perturbations that open leakage channels out of the local-support sector and thereby convert exact Zak-phase quantization into approximate robustness.

\subsection{Reflection-symmetric parent model}
\label{subsec:armchair_parent}

In the unit-cell basis
\begin{equation}
 \mathcal B=(A,B,C,D,E,F),
\end{equation}
the Bloch Hamiltonian of the simple six-site armchair nanoribbon is
\begin{equation}
 H_0(k)=
 \begin{pmatrix}
 \varepsilon&t&0&0&0&t\\
 t&\varepsilon&t&0&t_x\e^{-\ii k}&0\\
 0&t&\varepsilon&t&0&0\\
 0&0&t&\varepsilon&t&0\\
 0&t_x\e^{\ii k}&0&t&\varepsilon&t\\
 t&0&0&0&t&\varepsilon
 \end{pmatrix}.
 \label{eq:armchair_parent_hamiltonian}
\end{equation}
Here, $t$ denotes the intracell nearest-neighbor hopping and $t_x$ is the hopping across adjacent unit cells.  The real-space unit cell, the intercell bond, and the termination used for the open-chain calculation are displayed in the lattice inset of Fig.~\ref{fig:armchair_spectra} (b),(e), and (h) respectively for each simple, LSS, perturbated armchair nanoribbons.

The parent model is invariant under the left-right reflection that exchanges $A\leftrightarrow F$, $B\leftrightarrow E$, $C\leftrightarrow D$.
Denoting the corresponding permutation matrix by $M_{\mathrm{LR}}$, the symmetry condition is $M_{\mathrm{LR}}H_0(k)M_{\mathrm{LR}}^{-1}=H_0(-k)$.
The Hamiltonian has two momentum-independent eigenvectors,
\begin{align}
 v_+
 &=\frac{1}{2}
 \begin{pmatrix}
 1&0&-1&-1&0&1
 \end{pmatrix}^{T},
 \label{eq:armchair_even_flat_state}\\
 v_-
 &=\frac{1}{2}
 \begin{pmatrix}
 1&0&-1&1&0&-1
 \end{pmatrix}^{T},
 \label{eq:armchair_odd_flat_state}
\end{align}
corresponding to the two flat bands at $E_+ =\varepsilon+t$ and $E_- =\varepsilon-t$, respectively.
They have opposite reflection eigenvalues,
\begin{equation}
 M_{\mathrm{LR}}v_+=v_+,
 \qquad
 M_{\mathrm{LR}}v_-=-v_-.
 \label{eq:armchair_flat_state_parities}
\end{equation}
Because $v_\pm$ are independent of $k$ and belong to flat bands, their Fourier transforms are compact localized eigenstates confined to a single unit cell at $R_0$:
\begin{align}
 \ket{w_{+,R_0}}
 &=\frac{1}{2}
 \left(
 \ket{R_0,A}-\ket{R_0,C}-\ket{R_0,D}+\ket{R_0,F}
 \right),
 \label{eq:armchair_even_cls}\\
 \ket{w_{-,R_0}}
 &=\frac{1}{2}
 \left(
 \ket{R_0,A}-\ket{R_0,C}+\ket{R_0,D}-\ket{R_0,F}
 \right).
 \label{eq:armchair_odd_cls}
\end{align}
The amplitudes on the connector sites $B$ and $E$ vanish identically.
In addition to the two flat levels $E_\pm=\varepsilon\pm t$, the remaining four energies are
\begin{equation}
 E_{s,r}(k)
 =\varepsilon+s
 \sqrt{
 \frac{5t^2+t_x^2+r\sqrt{Q(k)}}{2}
 },
 \label{eq:armchair_dispersive_spectrum}
\end{equation}
where $Q(k)=\left(3t^2+t_x^2\right)^2+16t^3t_x\cos k$ and $s,r=\pm1$.

\begin{figure*}[!t]
    \centering
    \includegraphics[width=0.96\textwidth]{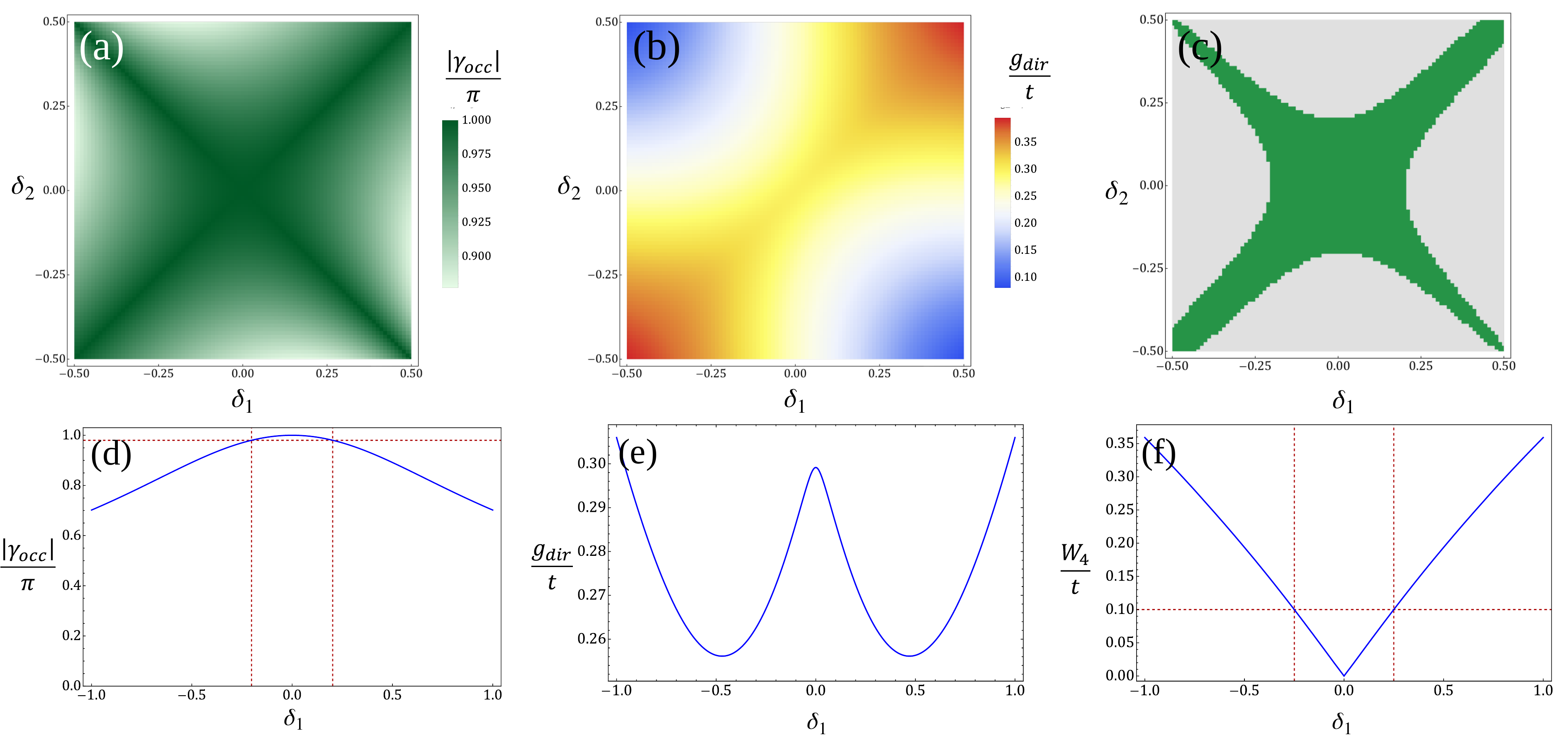}
    \caption{
Finite-perturbation response in the $(\delta_1,\delta_2)$ plane and perturbation response along the cut $\delta_2=0$.  
(a) Normalized occupied Zak phase, which remains closest to $\pi$ near $|\delta_1|\simeq|\delta_2|$.  
(b) Direct occupied gap, enhanced along $\delta_1\simeq\delta_2$ and reduced along $\delta_1\simeq-\delta_2$.  
(c) Robust region satisfying both $|\gamma_{\mathrm{occ}}|/\pi\geq0.98$ and $g_{\mathrm{dir}}/t\geq0.05$.  
(d) Zak phase of the occupied bands as a function of $\delta_1$, with the red dashed line marking the criterion $|\gamma_{\mathrm{occ}}|/\pi=0.98$.  
(e) The ratio between the direct gap $g_\mathrm{dir}$ between the fourth and fifth bands and $t$.  
(f) The ratio between the bandwidth of the LSS-derived fourth band, the nearly flat one, and $t$, with the red dashed line indicating $W_4/t=0.10$.  
    }
    \label{fig:fig5}
\end{figure*}

For the termination considered here, the occupied topological phase is most clearly isolated for $t_x>2t$.  Fig.~\ref{fig:armchair_spectra}(a)--(c) show a representative parent spectrum for $(\varepsilon,t,t_x)=(0,1,2.5)$.  The periodic spectrum contains the two analytic flat levels, whereas the open-boundary spectrum exhibits selected states inside the occupied bulk gap.  Their cell-resolved densities confirm localization near the ends of the finite chain.  For this parameter set, the six band Zak phases ordered from the lowest to the highest band are $(\gamma_1,\ldots,\gamma_6)
=(\pi,0,0,0,0,\pi)$, and the occupied-subspace Zak phase is $|\gamma_{\mathrm{occ}}|=\pi$.

Also, we investigated additional symmetry classes that yield quantized Zak phase in terms of hopping amplitudes and on-site energies for left-right reflection and $C_2$ symmetry cases in appendix \ref{app:additional_sym_6x6}.

\subsection{Ideal local-support-symmetry extension}
\label{subsec:armchair_ideal_lss}

We next add an auxiliary symmetry-breaking orbital $G$ with onsite energy $\varepsilon_G$ inside the hexagon, as shown in the inset of Fig.~\ref{fig:armchair_spectra}(e).  Following the general theory of the LSS, its coupling to the parent sector is chosen from the two known flat-band eigenvectors,
\begin{align}
h_{21}(k) = h_{12}(k)^\dag
 =g_1v_+^{T}+g_2v_-^{T}
 \label{eq:armchair_lss_coupling}
\end{align}
The resulting seven-site Hamiltonian is given by
\begin{equation}
 H_{\mathrm{LSS}}(k)=
 \begin{pmatrix}
 H_0(k)&h_{12}(k)\\
 h_{21}(k)&\varepsilon_G
 \end{pmatrix}.
 \label{eq:armchair_lss_hamiltonian}
\end{equation}
The coupling in Eq.~\eqref{eq:armchair_lss_coupling} generally breaks the global left--right reflection symmetry.  For example, $(g_1,g_2)=(1,1)$ couples $G$ only to the $A$--$C$ support, whereas $(g_1,g_2)=(1,-1)$ couples it only to the $D$--$F$ support.

The LSS mechanism becomes transparent by decomposing the parent Hilbert space into the two-dimensional flat-band subspace $\operatorname{span}\{v_+,v_-\}$ and its four-dimensional orthogonal complement.  Since $h_{12}$ lies entirely in the former subspace, every parent eigenvector $u_n(k)$ in the orthogonal complement satisfies
\begin{equation}
 h_{21}(k)\,u_n(k)=0.
 \label{eq:armchair_lss_kernel}
\end{equation}
Consequently,
\begin{equation}
 U_n(k)=
 \begin{pmatrix}
 u_n(k)\\0
 \end{pmatrix}
 \label{eq:armchair_extended_eigenstate}
\end{equation}
remains an exact eigenvector of $H_{\mathrm{LSS}}(k)$ with the same eigenvalue and Zak phase as in the parent model. The two parent flat-band states and the added orbital form an invariant three-dimensional subspace, while the remaining four parent eigenstates are exactly decoupled from the added orbital.  Thus the symmetry-breaking site can reorganize the flat levels without modifying the other four parent bands.  In particular, the topological occupied band retains its exactly quantized $\pi$ Zak phase even though the reflection symmetry is absent from the full seven-site Hamiltonian.

The decoupling of $U_n(k)$ from the auxiliary orbital can be understood directly in terms of destructive interference. Since the preserved parent eigenvector $u_n(k)$ belongs to the orthogonal complement of the two flat-band states, it satisfies $v_+^\dagger u_n(k)=v_-^\dagger u_n(k)=0.$ 
Using the explicit forms of $v_\pm$, these two conditions become $u_{n,A}-u_{n,C}-u_{n,D}+u_{n,F}=0$ and $u_{n,A}-u_{n,C}+u_{n,D}-u_{n,F}=0,$ and hence $u_{n,A}=u_{n,C}$ and $u_{n,D}=u_{n,F}$. 
The total amplitude transmitted from $u_n(k)$ to the auxiliary orbital $G$ is therefore $h_{21}(k)\,u_n(k) = \left[ (g_1+g_2) \bigl(u_{n,A}-u_{n,C}\bigr) + (g_2-g_1) \bigl(u_{n,D}-u_{n,F}\bigr) \right]/2 =0.$
Thus, the hopping amplitudes from the $A$ and $C$ sites cancel each other at $G$, as do those from the $D$ and $F$ sites. This destructive interference prevents the preserved state from acquiring any amplitude on the auxiliary orbital, thereby producing the exact extended eigenstate $U_n(k)$ in (\ref{eq:armchair_extended_eigenstate}). This cancellation is the real-space interpretation of the kernel condition $h_{21}(k)\,u_n(k)=0$.
This destructive interference is schematically drawn in Fig.~\ref{fig:fig4}(c).

Fig.~\ref{fig:armchair_spectra}(d)--(f) show the ideal LSS model for $(g_1,g_2,\varepsilon_G)=(1,1,-0.5)$, with the same parent parameters used above.  The comparison with Figs.~\ref{fig:armchair_spectra}(a)--\ref{fig:armchair_spectra}(c) shows that the reflection-breaking extension preserves the relevant parent dispersive bands and their boundary response while reconstructing the three-level LSS sector. The occupied-subspace Zak phase remains exactly $|\gamma_{\mathrm{occ}}|/\pi=1$, and the selected open-boundary states remain localized at the ends of the finite chain. The relation between $g_1$, $g_2$, and $\varepsilon_G$ can be interpreted as a hyperbolic equation in the three-dimensional space. We derived the function that shows relation of these variables in appendix \ref{app:hyperbolic_gap_armchair}. The function indicates boundary that determines possible gap opening region to maintain gap-opening in $\kappa_1$ and $\kappa_2$ plane where $\kappa_1$ and $\kappa_2$ are dimensionless variables of $g_1$ and $g_2$. Detailed explanation is listed in Appendix \ref{app:hyperbolic_gap_armchair}.

\subsection{Leakage perturbations and near-quantized response}
\label{subsec:armchair_perturbation}

To model deviations from the ideal coupling geometry satisfying destructive interference, we introduce additional hoppings $t_1$ and $t_2$ between the auxiliary site $G$ and the connector sites $B$ and $E$, respectively.

In contrast to the coupling in Eq.~\eqref{eq:armchair_lss_coupling}, these perturbation terms couple the auxiliary site directly to the connector sites, where the ideal flat-sector eigenstates have vanishing amplitudes. The additional hoppings inject amplitudes $t_1\psi_G$ and $t_2\psi_G$ into $B$ and $E$, respectively, thereby coupling the flat-band sector to the remaining states. Consequently, the compact localized eigenstates originating from Eqs.~\eqref{eq:armchair_even_flat_state} and \eqref{eq:armchair_odd_flat_state} are generally no longer exact eigenstates of the perturbed Hamiltonian. Fig.~\ref{fig:fig4}(b) illustrates this process for $t_1\neq0$ and $t_2=0$. The corresponding bands can therefore acquire finite dispersion. Fig.~\ref{fig:armchair_spectra}(g)--(i) present a representative perturbed model with $(t_1,t_2)=(0.2,0)$. The weak dispersion of the band derived from the ideal flat-band sector is highlighted in the inset of Fig.~\ref{fig:armchair_spectra}(g), while the open-chain spectrum and the real-space density in Figs.~\ref{fig:armchair_spectra}(h) and \ref{fig:armchair_spectra}(i) show that the in-gap boundary states persist under this finite perturbation.

The perturbations affect not only the compact localization of the flat-band sector but also the exact Zak-phase quantization inherited through LSS. In the ideal construction, the preserved parent eigenstates $u_n(k)$ satisfy the kernel condition in Eq.~\eqref{eq:armchair_lss_kernel}. However, the additional hoppings open leakage channels from the parent sector into the auxiliary site $G$. Denoting the ideal coupling by $h_{21}^{(0)}(k)$, the perturbed coupling can be written as
\begin{equation}
    h_{21}^{\mathrm{pert}}(k)
    =h_{21}^{(0)}(k)
    +t_1\langle B|+t_2\langle E|.
    \label{eq:armchair_perturbed_coupling}
\end{equation}
Applying this coupling to the preserved parent eigenstate gives
\begin{equation}
    h_{21}^{\mathrm{pert}}(k)u_n(k)
    =t_1u_{n,B}(k)+t_2u_{n,E}(k),
    \label{eq:armchair_lss_kernel_perturbed}
\end{equation}
which is generally nonzero.

Since the right-hand side of Eq.\eqref{eq:armchair_lss_kernel_perturbed} is generally nonzero, the embedded state in Eq.\eqref{eq:armchair_extended_eigenstate} is no longer an exact eigenstate of the perturbed Hamiltonian. As a result, the parent-derived states are no longer guaranteed to remain exact eigenstates of the perturbed Hamiltonian, and their Zak phases are no longer constrained to their ideal quantized values by LSS. For the representative parameter set, the occupied-subspace Zak phase changes from $|\gamma_{\mathrm{occ}}|/\pi=1$ in the ideal model to approximately $0.981$. The persistence of a value close to $\pi$ therefore reflects perturbative robustness rather than exact LSS protection.

For a seven-band model, we consider the filling at which the four lowest bands are occupied.
Then, we investigate the direct gap $g_\mathrm{dir}$ between the lowest unoccupied band($E_5(k)$) and the highest occupied band ($E_4(k)$) and the bandwidth $W_4$ of $E_4(k)$.
A finite $g_{\mathrm{dir}}$ keeps the four-dimensional occupied subspace isolated, whereas $W_4$ quantifies the perturbation-induced loss of flatness and is not itself a topological criterion.  
Introducing the dimensionless leakage strengths
\begin{equation}
 \delta_1=\frac{t_1}{t},
 \qquad
 \delta_2=\frac{t_2}{t},
 \label{eq:armchair_dimensionless_perturbations}
\end{equation}
Fig.~\ref{fig:fig5} (d)--(f) shows the response along the cut $\delta_2=0$.  Starting from the ideal point $\delta_1=0$, the occupied Zak phase departs continuously from $\pi$ without an accompanying closure of the direct occupied gap.  For the parameters used in the figure, $|\gamma_{\mathrm{occ}}|/\pi$ remains above the operational threshold $0.98$ for $|\delta_1|<0.20231$.  The fourth-band bandwidth grows with $|\delta_1|$ and reaches the auxiliary flatness threshold $W_4/t=0.10$ at approximately $\delta_1=\pm0.24962$.

The two-dimensional maps in Fig.~\ref{fig:fig5} (a)--(c) extend this analysis to the full $(\delta_1,\delta_2)$ plane.  The Zak phase remains closest to $\pi$ near the diagonals $|\delta_1|\simeq|\delta_2|$, whereas the direct gap depends strongly on the relative sign of the two leakage amplitudes: it is enhanced along $\delta_1\simeq\delta_2$ and reduced along $\delta_1\simeq-\delta_2$.  The common robust region shown in Fig.~\ref{fig:fig5}(c) is defined by the simultaneous numerical criteria
\begin{equation}
 \frac{|\gamma_{\mathrm{occ}}|}{\pi}\geq0.98,
 \qquad
 \frac{g_{\mathrm{dir}}}{t}\geq0.05.
 \label{eq:armchair_robustness_criteria}
\end{equation}
This finite region demonstrates that the boundary response and a near-$\pi$ occupied Zak phase can remain stable beyond the exactly solvable LSS limit.  Nevertheless, only the ideal kernel condition produces exact quantization in the globally reflection-broken system.

\section{Conclusion}
\label{sec:conclusion}

In summary, we have established a mechanism for exact Zak-phase
quantization in one-dimensional systems whose full Bloch Hamiltonian does
not possess the corresponding protecting crystalline symmetry. Starting
from a parent insulator with a symmetry-quantized Zak phase, we introduced
additional symmetry-breaking degrees of freedom while constraining their
couplings such that destructive interference prevents the relevant Bloch
states from leaking into the added sector. The protected states therefore
retain exactly the same wave functions as their parent counterparts on
their nonzero support and inherit their Zak phase unchanged. We
demonstrated this mechanism in a minimal inversion-broken model and in an
armchair-nanoribbon model with globally broken reflection symmetry. In
both cases, the exact $\pi$ Zak phase is accompanied by boundary-localized
states. We further showed that perturbations violating the exact
destructive-interference condition open leakage channels into the
symmetry-breaking sector, continuously dequantizing the Zak phase while
leaving a finite regime in which a near-$\pi$ geometric phase and the
associated boundary response remain robust.

An important implication of our results concerns the relation between LSS
and previously studied mechanisms for Zak-phase quantization in systems
where an apparent protecting symmetry is broken. In some cases, breaking
one conventional symmetry does not destroy the quantization because another
global symmetry remains operative. For example, a quantized Zak phase can
survive the breaking of chiral symmetry when inversion symmetry is
preserved~\cite{jiao2021experimentally}. More closely related to the present
work are systems in which the protecting symmetry is absent from the full
Hamiltonian but survives in a hidden form. Hidden-inversion constructions
based on coupled SSH systems, for instance, possess a decoupled topological
subspace whose effective Hamiltonian retains inversion symmetry despite the
absence of inversion symmetry in the full system~\cite{xiao2016coexistence}.
More generally, latent symmetries can emerge in isospectral effective
Hamiltonians obtained by subsystem reduction~\cite{rontgen2021latent}, and
this principle has recently been used to construct SSH-like systems with
neither manifest reflection nor chiral symmetry but with a hidden reflection
symmetry that quantizes the Zak phase and protects topological edge
states~\cite{rontgen2024topological}. Related approaches have also shown that
boundary states may survive under symmetry-breaking perturbations that
preserve only an appropriate sub-symmetry~\cite{wang2023subsymmetry}.
These works demonstrate that the apparent absence of a protecting symmetry
from the full Hamiltonian does not necessarily imply its complete loss:
the relevant symmetry may remain globally in another form or reappear in a
reduced sector of the Hilbert space.

The minimal LSS construction considered here is distinct from such a
hidden-symmetry block-reduction mechanism. As shown in
Appendix~\ref{app:no_hidden_symmetry}, no nontrivial $k$-independent
invariant block containing the protected band exists in the enlarged
three-band Hamiltonian. The destructive-interference condition instead
selects a momentum-dependent kernel that follows the protected Bloch state
throughout the Brillouin zone. Consequently, there is no fixed
symmetry-preserving topological block to which the exact $\pi$ Zak phase
can be reassigned. The quantization is inherited directly from the
symmetry-constrained parent eigenstate, which remains an exact eigenstate
of the enlarged, globally inversion-broken Hamiltonian. In this sense,
LSS provides a state-selective protection mechanism that goes beyond the
hidden-symmetry block construction of Ref.~\cite{xiao2016coexistence}.

The armchair model also reveals when these two viewpoints become
equivalent. As demonstrated explicitly in
Appendix~\ref{app:armchair_hidden}, its destructive-interference condition
leaves a $k$-independent four-dimensional subspace exactly decoupled from
the symmetry-breaking sector. A momentum-independent basis transformation
then decomposes the full Hamiltonian into a reflection-symmetric
topological block and a complementary block in which reflection is
generically broken. The exact Zak-phase quantization in this particular
realization therefore admits a hidden-reflection interpretation analogous
to the hidden-subspace mechanism discussed in
Refs.~\cite{xiao2016coexistence,rontgen2024topological}. This comparison
shows that hidden-symmetry protection can emerge as a special realization
of LSS when the kernel of the auxiliary symmetry-breaking coupling contains a common
momentum-independent invariant subspace. LSS itself requires only the
weaker, momentum-resolved condition that the symmetry-breaking coupling
annihilate the protected Bloch states.

This distinction suggests a broader classification of local-support
protection according to the momentum dependence of the decoupled
subspace. When destructive interference produces a fixed invariant
subspace, the resulting topology may be recast in terms of a hidden
symmetry of an effective block, as in the armchair example. When the
kernel instead follows a momentum-dependent Bloch subspace, as in the
minimal model, no such fixed hidden-symmetry block need exist. The latter
possibility enlarges the conventional notion of symmetry protection:
quantized quantum geometry can survive not because a protecting symmetry
is recoverable in a fixed reduced block, but because destructive
interference prevents the symmetry-breaking degrees of freedom from
altering the wave functions that carry the geometric phase. This
support-based viewpoint provides a general design principle for
topological states in enlarged and globally symmetry-broken systems and
may be applicable to other symmetry-quantized geometric and topological
quantities beyond the Zak phase.

\begin{acknowledgments}
J.W.R. was supported by the National Research Foundation
of Korea (NRF) grants funded by the Korean government
(MSIT) (Grant Nos. RS-2023-NR068116 and RS-2026-25597121)
and the Ministry of Education
(Grant No. RS-2023-00285390).
\end{acknowledgments}

\appendix

\section{Direct Zak-Phase Calculation for the Two-Band Model}
\label{app:direct_phase_cal}

\subsection{Brillouin-zone trajectories and winding}

The complex projection
\[
\Delta(k)=d_x(k)+i d_y(k)
\]
provides a geometric diagnostic of the two parent branches.  In the perfect Pythagorean case, it
traces an ellipse centered at the origin, whereas in the general/imperfect Pythagorean case, the ellipse is
shifted by $b_0$ along the real axis.  The occupied band is in the nontrivial
gapped regime when the corresponding closed curve encloses the origin once.
For the shifted ellipse this gives $|a_0|>|b_0|$; equality marks the
gap-closing boundary.

In this appendix, we show why the Zak phase of the nontrivial two-band flat-band
branches is quantized to
\[
\gamma=\pi \quad (\mathrm{mod}\ 2\pi).
\]
The proof is based on the winding of the complex function
\[
\Delta(k)=d_x(k)+i d_y(k)
\]
and on the parity structure imposed by the inversion symmetry.

We consider a two-band Bloch Hamiltonian
\begin{align*}
H(k)
={}&d_0(k)I+d_x(k)\sigma_x
\\
&+d_y(k)\sigma_y+d_z(k)\sigma_z.
\end{align*}
The scalar term \(d_0(k)I\) only shifts the energy and does not change the
eigenvectors. Therefore, the Zak phase is determined by the vector
\[
\mathbf d(k)=(d_x(k),d_y(k),d_z(k)).
\]

Let
\[
\Delta(k)=d_x(k)+i d_y(k),\qquad s(k)=d_z(k).
\]
We write
\[
\Delta(k)=|\Delta(k)|e^{i\phi(k)}
\]
and introduce \(\theta(k)\) by
\[
\cos\theta(k)=\frac{s(k)}{|\mathbf d(k)|},\qquad
\sin\theta(k)=\frac{|\Delta(k)|}{|\mathbf d(k)|}.
\]
A normalized eigenvector for the lower band may be chosen as
\[
|u(k)\rangle=
\begin{pmatrix}
\sin\frac{\theta(k)}{2}e^{-i\phi(k)}\\
-\cos\frac{\theta(k)}{2}
\end{pmatrix}.
\]
Then the Berry connection is
\[
A(k)=i\langle u(k)|\partial_k u(k)\rangle
=
\sin^2\frac{\theta(k)}{2}\,\partial_k\phi(k).
\]
Using
\[
\sin^2\frac{\theta(k)}{2}
=
\frac{1-\cos\theta(k)}{2},
\]
the Zak phase becomes
\begin{align*}
\gamma
&=
\int_{-\pi}^{\pi} A(k)\,dk
\\
&=
\frac{1}{2}\int_{-\pi}^{\pi}\partial_k\phi(k)\,dk
-\frac{1}{2}\int_{-\pi}^{\pi}
\cos\theta(k)\partial_k\phi(k)\,dk.
\end{align*}
Thus the Zak phase is separated into a winding contribution and an additional
\(d_z(k)\)-dependent contribution.

\subsection{Case 1: Perfect Pythagorean Case}

For the first nontrivial branch, the Hamiltonian components are
\begin{align*}
d_0(k)=a_0, \quad
&d_x(k)=\eta a_0\cos k, 
\\
d_y(k)=c_1\sin k, \quad
&d_z(k)=d_1\sin k,
\end{align*}
where \(\eta=\pm1\), together with the flat-band condition
\[
a_0^2=c_1^2+d_1^2.
\]
Then
\[
\Delta(k)=\eta a_0\cos k+i c_1\sin k,
\qquad
s(k)=d_1\sin k.
\]
The magnitude of \(\mathbf d(k)\) is
\[
|\mathbf d(k)|
=
\sqrt{a_0^2\cos^2k+(c_1^2+d_1^2)\sin^2k}
=
|a_0|.
\]
Therefore the band is flat.

The curve \(\Delta(k)\) forms an ellipse centered at the origin in the complex plane.
As \(k\) goes through the Brillouin zone, this ellipse encloses the origin once, provided
that the bulk gap remains open. Therefore the phase \(\phi(k)\) winds by
\[
\phi(\pi)-\phi(-\pi)=2\pi W,
\]
where \(W=\pm1\) is the winding number. Hence the winding contribution to the Zak
phase is
\[
\frac{1}{2}\int_{-\pi}^{\pi}\partial_k\phi(k)\,dk
=
\pi W
=
\pi
\quad
(\mathrm{mod}\ 2\pi).
\]

It remains to check the second term. From the definition of \(\theta(k)\),
\[
\cos\theta(k)
=
\frac{d_1\sin k}{|a_0|}.
\]
This is an odd function of \(k\). On the other hand, since
\[
\Delta(-k)=\Delta(k)^*,
\]
we have
\[
\phi(-k)=-\phi(k)\quad(\mathrm{mod}\ 2\pi),
\]
and therefore
\[
\partial_k\phi(-k)=\partial_k\phi(k).
\]
Thus \(\partial_k\phi(k)\) is an even function of \(k\). Hence
\[
\cos\theta(k)\partial_k\phi(k)
\]
is odd, and its integral over the symmetric Brillouin zone vanishes:
\[
\int_{-\pi}^{\pi}\cos\theta(k)\partial_k\phi(k)\,dk=0.
\]
Consequently,
\[
\gamma=\pi\quad(\mathrm{mod}\ 2\pi).
\]

\subsection{Case 2: Imperfect Pythagorean Case}

For the second nontrivial branch, the Hamiltonian components are
\begin{align*}
d_0(k)=a_0+\eta b_0\cos k, \quad &d_x(k)=b_0+\eta a_0\cos k,
\\
d_y(k)=c_1\sin k, \quad
&d_z(k)=d_1\sin k,
\end{align*}

where \(\eta=\pm1\), together with the flat-band condition
\[
a_0^2-b_0^2=c_1^2+d_1^2.
\]
The corresponding complex function is
\[
\Delta(k)=b_0+\eta a_0\cos k+i c_1\sin k,
\qquad
s(k)=d_1\sin k.
\]
Then
\[
|\mathbf d(k)|
=
\sqrt{(b_0+\eta a_0\cos k)^2+(c_1^2+d_1^2)\sin^2k}.
\]
Using the flat-band condition,
\[
|\mathbf d(k)|
=
\sqrt{(b_0+\eta a_0\cos k)^2+(a_0^2-b_0^2)\sin^2k}.
\]
Expanding the expression inside the square root gives
\[
|\mathbf d(k)|
=
\sqrt{(a_0+\eta b_0\cos k)^2}.
\]
Hence
\[
|\mathbf d(k)|=|a_0+\eta b_0\cos k|.
\]
For the flat band to be separated from the other band, the bulk gap must remain open.
In particular, the nontrivial gapped regime is
\[
|a_0|>|b_0|.
\]
The equality case \(|a_0|=|b_0|\) corresponds to a gap-closing point and should be
excluded when defining the Zak phase.

In this case,
\[
\Delta(k)=b_0+\eta a_0\cos k+i c_1\sin k
\]
forms an ellipse shifted along the real axis. The origin lies inside this ellipse if
\[
|a_0|>|b_0|.
\]
Therefore \(\Delta(k)\) winds around the origin once, and
\[
\phi(\pi)-\phi(-\pi)=2\pi W,\qquad W=\pm1.
\]
The winding contribution is then
\[
\frac{1}{2}\int_{-\pi}^{\pi}\partial_k\phi(k)\,dk
=
\pi W
=
\pi
\quad
(\mathrm{mod}\ 2\pi).
\]

The remaining term again vanishes by parity. Since
\[
\cos\theta(k)
=
\frac{s(k)}{|\mathbf d(k)|}
=
\frac{d_1\sin k}{|a_0+\eta b_0\cos k|},
\]
\(\cos\theta(k)\) is an odd function of \(k\). Also,
\[
\Delta(-k)=\Delta(k)^*,
\]
so \(\phi(k)\) is odd up to \(2\pi\), and \(\partial_k\phi(k)\) is even. Therefore
\[
\cos\theta(k)\partial_k\phi(k)
\]
is odd, and
\[
\int_{-\pi}^{\pi}\cos\theta(k)\partial_k\phi(k)\,dk=0.
\]
Thus,
\[
\gamma=\pi\quad(\mathrm{mod}\ 2\pi).
\]

The winding argument above assumes $c_1\neq0$, for which
$\Delta(k)=d_x(k)+i d_y(k)$ traces a nondegenerate ellipse. However, the case $c_1=0$ does not represent a physical singularity. From the flat-band condition $c_1^2+d_1^2=a_0^2-b_0^2>0$, we have $d_1\neq 0$ when $c_1=0$. By continuously varying $c_1$ and $d_1$ while keeping $c_1^2+d_1^2$ fixed, the system can be deformed from $c_1\neq 0$ to $c_1=0$ without closing the bulk gap or breaking inversion symmetry. Since the Zak phase remains quantized under inversion symmetry and cannot change continuously without a gap closing, the result $\gamma=\pi$ $\pmod {2\pi}$ also holds for $c_1=0$. Thus, the apparent singularity arises only from the choice of the complex projection $\Delta(k)$, rather than from the physical band structure.

\section{Additional symmetry classes}
\label{app:additional_sym_6x6}
\subsection{Left--right reflection symmetry of the armchair nanoribbon}

We first consider a generalized six-site armchair nanoribbon with
left--right reflection symmetry,
\[
A\leftrightarrow F,
\qquad
B\leftrightarrow E,
\qquad
C\leftrightarrow D.
\]
The reflection operation is accompanied by the momentum inversion
$k\rightarrow -k$.
In the basis
\[
\begin{aligned}
\mathcal{B}=\{&
\lvert A\rangle,\lvert B\rangle,\lvert C\rangle,\lvert D\rangle,\lvert E\rangle,\lvert F\rangle
\},
\end{aligned}
\]
the Bloch Hamiltonian of the model is
\[
H(k)=
\begin{pmatrix}
\varepsilon_A & t_{ab} & 0 & 0 & 0 & t_{fa} \\
t_{ab} & \varepsilon_B & t_{bc} & 0 & t_xe^{-ik} & 0 \\
0 & t_{bc} & \varepsilon_C & t_{cd} & 0 & 0 \\
0 & 0 & t_{cd} & \varepsilon_C & t_{bc} & 0 \\
0 & t_xe^{ik} & 0 & t_{bc} & \varepsilon_B & t_{ab} \\
t_{fa} & 0 & 0 & 0 & t_{ab} & \varepsilon_A
\end{pmatrix}.
\]
The matrix representation of the mirror operation is
\[
M_{LR}
=
\begin{pmatrix}
0 & J_{3}\\
J_{3} & 0
\end{pmatrix},
\]
which satisfies
\[
M_{LR}H(k)M_{LR}^{-1}=H(-k),
\]
where $J_3$ is
\[
J_3=\begin{pmatrix}
    0 & 0 & 1 \\
    0 & 1 & 0 \\
    1 & 0 & 0
\end{pmatrix}.
\]

To exploit the left--right reflection symmetry, we introduce the
symmetric and antisymmetric basis states
\begin{align*}
\lvert A_{\pm}\rangle
&=
\frac{\lvert A\rangle\pm\lvert F\rangle}{\sqrt{2}},\\
\lvert B_{\pm}\rangle
&=
\frac{\lvert B\rangle\pm\lvert E\rangle}{\sqrt{2}},\\
\lvert C_{\pm}\rangle
&=
\frac{\lvert C\rangle\pm\lvert D\rangle}{\sqrt{2}}.
\end{align*}
The corresponding symmetry-adapted basis is
\[
\begin{aligned}
\mathcal{B}'=\{&
\lvert A_+\rangle,\lvert B_+\rangle,\lvert C_+\rangle,\lvert A_-\rangle,\lvert B_-\rangle,\lvert C_-\rangle
\}.
\end{aligned}
\]
The unitary transformation matrix is
\[
U=
\begin{pmatrix}
\frac{1}{\sqrt{2}} & 0 & 0 & \frac{1}{\sqrt{2}} & 0 & 0 \\
0 & \frac{1}{\sqrt{2}} & 0 & 0 & \frac{1}{\sqrt{2}} & 0 \\
0 & 0 & \frac{1}{\sqrt{2}} & 0 & 0 & \frac{1}{\sqrt{2}} \\
0 & 0 & \frac{1}{\sqrt{2}} & 0 & 0 & -\frac{1}{\sqrt{2}} \\
0 & \frac{1}{\sqrt{2}} & 0 & 0 & -\frac{1}{\sqrt{2}} & 0 \\
\frac{1}{\sqrt{2}} & 0 & 0 & -\frac{1}{\sqrt{2}} & 0 & 0
\end{pmatrix}.
\]
By applying this unitary transformation,
\[
\widetilde{H}(k)=U^\dagger H(k)U,
\]
the Hamiltonian becomes
\[
\widetilde{H}(k)=
\begin{pmatrix}
a_+ & t_{ab} & 0 & 0 & 0 & 0 \\
t_{ab} & \varepsilon_B+t_x\cos k & t_{bc}
       & 0 & it_x\sin k & 0 \\
0 & t_{bc} & c_+ & 0 & 0 & 0 \\
0 & 0 & 0 & a_- & t_{ab} & 0 \\
0 & -it_x\sin k & 0 & t_{ab}
  & \varepsilon_B-t_x\cos k & t_{bc} \\
0 & 0 & 0 & 0 & t_{bc} & c_-
\end{pmatrix},
\]
where
\[
a_\pm=\varepsilon_A\pm t_{fa},
\qquad
c_\pm=\varepsilon_C\pm t_{cd}.
\]
For a flat band to exist, given $\tilde{H}(k)$ has to satisfy
\[
\det [\tilde{H}(k)-E_f I_6]=0 \; \forall k,
\]
where $I_6$ is $6\times6$ identity matrix. From the determinant above, one can obtain following result:
\begin{align*}
    &\det[\tilde{H}(k)-E_f I_6] \\
    &=A_+A_-C_+C_-[\D_+\D_--t_x^2+t_x(\D_--\D_+)\cos{k}],
\end{align*}
where
\begin{align*}
    A_\pm=a_\pm-E_f, \quad B=\varepsilon_B-E_f,\quad C_\pm=c_\pm-E_f.
\end{align*}
For $t_x\neq0$ and $A_\pm,\;C_\pm\neq0$, the flat-band condition is
\[
\Delta_+(E_f)=\Delta_-(E_f),
\qquad
\Delta_\pm^{\,2}(E_f)=t_x^2,
\]
where
\begin{align*}
\Sigma_\pm(E_f)
&=
\frac{t_{ab}^{\,2}}{A_\pm}
+
\frac{t_{bc}^{\,2}}{C_\pm},\\
\Delta_\pm(E_f)
&=
B-\Sigma_\pm(E_f).
\end{align*}

Equivalently,
\[
\frac{t_{ab}^{\,2}}{a_+-E_f}
+
\frac{t_{bc}^{\,2}}{c_+-E_f}
=
\frac{t_{ab}^{\,2}}{a_--E_f}
+
\frac{t_{bc}^{\,2}}{c_--E_f}.
\]

Under this condition, the flat-band energy satisfies
\[
E_f
=
\varepsilon_B
-
\Sigma_\pm(E_f)
-
s t_x,
\qquad
s=\pm1
\]

The topology of the resulting flat band can be determined from its reflection eigenvalues at the inversion-invariant momenta $k_\star=0,\pi$. At these momenta, the intersector coupling proportional to $\sin k$ vanishes, and the transformed Hamiltonian separates into the symmetric and antisymmetric sectors $H_\pm(k_\star)$.
At these two symmetry points, $\tilde{H}(k)$ is written as

\[
\tilde{H}(k_*)=
\begin{pmatrix}
    H_+(k_*) & 0 \\
    0 & H_-(k_*)
\end{pmatrix}, \quad k_*=0,\pi
\]

Their characteristic determinants at the flat-band energy are
\begin{align*}
\det[H_+(k_\star)-E_f I_3]
&=A_+C_+t_x(s+\cos k_\star),\\
\det[H_-(k_\star)-E_f I_3]
&=A_-C_-t_x(s-\cos k_\star).
\end{align*}
\

For the regular flat-band branch, the corresponding reflection eigenvalues are therefore
\begin{align*}
    \xi(0)=-s,\qquad \xi(\pi)=s,
    \qquad s=\pm1.
\end{align*}

Provided that the flat band remains spectrally isolated, the Zak phase satisfies
\begin{align*}
    e^{i\gamma_{\mathrm{flat}}}
    =\xi(0)\xi(\pi)=-1.
\end{align*}
Hence,
\begin{align*}
    \gamma_{\mathrm{flat}}=\pi \quad (\mathrm{mod}\ 2\pi).
\end{align*}

\subsection{\texorpdfstring{$C_2$}{C2}-symmetric armchair nanoribbon}

We next consider a generalized armchair nanoribbon with twofold rotational
symmetry denoted by $C_2$.
The symmetry operation exchanges the lattice sites according to
\[
A\leftrightarrow D,
\qquad
B\leftrightarrow E,
\qquad
C\leftrightarrow F,
\]
and is accompanied by the momentum inversion $k\rightarrow -k$.
In the basis
\[
\begin{aligned}
\mathcal{B}=\{&
\lvert A\rangle,\lvert B\rangle,\lvert C\rangle,\lvert D\rangle,\lvert E\rangle,\lvert F\rangle
\},
\end{aligned}
\]
the Bloch Hamiltonian is
\[
H(k)=
\begin{pmatrix}
\varepsilon_A & t_{ab} & 0 & 0 & 0 & t_{cd} \\
t_{ab} & \varepsilon_B & t_{bc} & 0 & t_x e^{-ik} & 0 \\
0 & t_{bc} & \varepsilon_C & t_{cd} & 0 & 0 \\
0 & 0 & t_{cd} & \varepsilon_A & t_{ab} & 0 \\
0 & t_x e^{ik} & 0 & t_{ab} & \varepsilon_B & t_{bc} \\
t_{cd} & 0 & 0 & 0 & t_{bc} & \varepsilon_C
\end{pmatrix}.
\]
The matrix representation of the $C_2$ operation is
\[
M_{C_2}
=
\begin{pmatrix}
0 & I_{3}\\
I_{3} & 0
\end{pmatrix},
\]
which satisfies
\[
M_{C_2}H(k)M_{C_2}^{-1}=H(-k).
\]

To construct a symmetry-adapted basis, we introduce the symmetric
and antisymmetric basis states
\begin{align*}
\lvert A_{\pm}\rangle
&=
\frac{\lvert A\rangle\pm\lvert D\rangle}{\sqrt{2}},\\
\lvert B_{\pm}\rangle
&=
\frac{\lvert B\rangle\pm\lvert E\rangle}{\sqrt{2}},\\
\lvert C_{\pm}\rangle
&=
\frac{\lvert C\rangle\pm\lvert F\rangle}{\sqrt{2}}.
\end{align*}
The corresponding basis is
\[
\begin{aligned}
\mathcal{B}'=\{&
\lvert A_+\rangle,\lvert B_+\rangle,\lvert C_+\rangle,\\[-2pt]
&
\lvert A_-\rangle,\lvert B_-\rangle,\lvert C_-\rangle
\}.
\end{aligned}
\]
The unitary transformation matrix is
\[
U=
\begin{pmatrix}
\frac{1}{\sqrt{2}} & 0 & 0
    & \frac{1}{\sqrt{2}} & 0 & 0 \\
0 & \frac{1}{\sqrt{2}} & 0
    & 0 & \frac{1}{\sqrt{2}} & 0 \\
0 & 0 & \frac{1}{\sqrt{2}}
    & 0 & 0 & \frac{1}{\sqrt{2}} \\
\frac{1}{\sqrt{2}} & 0 & 0
    & -\frac{1}{\sqrt{2}} & 0 & 0 \\
0 & \frac{1}{\sqrt{2}} & 0
    & 0 & -\frac{1}{\sqrt{2}} & 0 \\
0 & 0 & \frac{1}{\sqrt{2}}
    & 0 & 0 & -\frac{1}{\sqrt{2}}
\end{pmatrix}.
\]

By applying the unitary transformation
\[
\widetilde{H}(k)=U^\dagger H(k)U,
\]
the Hamiltonian becomes

\[
\widetilde{H}(k)=
\begin{pmatrix}
\varepsilon_A & t_{ab} & t_{cd} & 0 & 0 & 0 \\
t_{ab} & \varepsilon_B+t_x\cos k & t_{bc}
& 0 & i t_x\sin k & 0 \\
t_{cd} & t_{bc} & \varepsilon_C & 0 & 0 & 0 \\
0 & 0 & 0 & \varepsilon_A & t_{ab} & -t_{cd} \\
0 & -i t_x\sin k & 0 & t_{ab}
& \varepsilon_B-t_x\cos k & t_{bc} \\
0 & 0 & 0 & -t_{cd} & t_{bc} & \varepsilon_C
\end{pmatrix}.
\]
For flat band condition, given $\tilde{H}(k)$ has to satisfy
\[
\det [\tilde{H}(k)-E_f I_6]=0 \;\forall k,
\]
where $I_6$ is $6\times6$ identity matrix. From the determinant above, one can obtain following result:
\begin{align*}
    &\det[\tilde{H}(k)-E_f I_6] \\
    &=(BD-K)^2-T^2-D^2t_x^2-2DTt_x\cos{k},
\end{align*}
where
\begin{align*}
    &A=\varepsilon_A-E_f, \quad B=\varepsilon_B-E_f, \quad C=\varepsilon_C-E_f  \\
    &D=AC-t_{cd}^2, \quad K=Ct_{ab}^2+At_{bc}^2, \quad
    T=2t_{ab}t_{bc}t_{cd}.
\end{align*}
For $t_x\neq0$, this condition reduces to
\[
-2DTt_x=0 \;\text{and}\;(BD-K)^2-T^2-D^2t_x^2=0.
\]
Since $t_{ab},t_{bc},t_{cd}\neq0$ in this lattice, $D$ has to be zero. Thus, flat band conditions are
\[
\begin{cases}
    (\varepsilon_A-E_f)(\varepsilon_C-E_f)=t_{cd}^2 \\
    (\varepsilon_C-E_f)t_{ab}^2=(\varepsilon_A-E_f)t_{bc}^2
\end{cases}
\]
from $D=0$ and substitute $D=0$ into second flat band condition, respectively.
Under this condition, the flat-band energy satisfies
\[
E_f=\varepsilon_A\mp\frac{t_{ab}t_{cd}}{t_{bc}}=\varepsilon_C\mp\frac{t_{bc}t_{cd}}{t_{ab}}.
\]
By defining $s=\pm1$, $E_f=\varepsilon_A\mp\frac{t_{ab}t_{cd}}{t_{bc}}=\varepsilon_C\mp\frac{t_{bc}t_{cd}}{t_{ab}}$ can be written as
\[
\varepsilon_A-E_f=s\frac{t_{ab}t_{cd}}{t_{bc}}, \quad \varepsilon_C-E_f=s\frac{t_{bc}t_{cd}}{t_{ab}}.
\]
This yields the eigenvector of $H(k)$ to be
\[
|v_s\rangle=\frac{1}{\sqrt{2(t_{ab}^2+t_{bc}^2)}}
    \begin{pmatrix}
        t_{bc}\\
        0\\
        -t_{ab}\\
        st_{bc}\\
        0\\
        -st_{ab}
    \end{pmatrix}.
\]
Since the normalized eigenvector is independent of $k$, its Berry connection vanishes identically in the chosen periodic Bloch basis. Therefore, the corresponding isolated flat band has a Zak phase of $\gamma_{\mathrm{flat}}=0$ (mod $2\pi$).






\begin{figure*}[!t]
    \centering
    \includegraphics[width=0.96\textwidth]{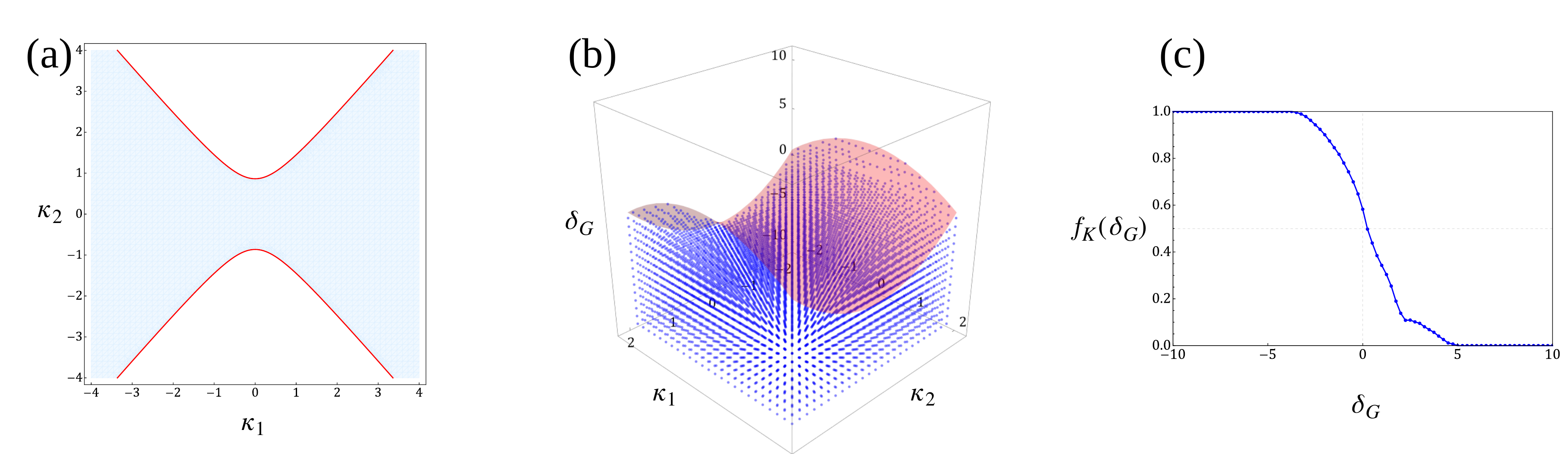}
    \caption{
    Ideal-LSS gap-closing geometry and finite-parameter robustness of the armchair nanoribbon model.
    (a) Analytic gap-closing boundary in the \((\kappa_1,\kappa_2)\) plane at fixed detuning \(\delta_G=-0.5\), obtained from Eq.~\eqref{eq:appC_hyperbola}. The red curves form a hyperbola separating parameter regions with different direct-gap status.
    (b) Gap-closing surface \(\delta_G(\kappa_1,\kappa_2)\) given by Eq.~\eqref{eq:appC_gap_closing_surface}, displaying the saddle-shaped dependence of the critical auxiliary-site detuning on the two LSS couplings. The surface passes through \((\kappa_1,\kappa_2,\delta_G)=(0,0,e_c)\).
    (c) Robust fraction \(f_K(\delta_G)\), defined in Eqs.~\eqref{eq:appC_robust_fraction} and ~\eqref{eq:appC_robust_indicator}, evaluated over the sampled coupling domain \(\mathcal{D}_K=[-K,K]^2\). The curve gives the fraction of sampled coupling parameters that simultaneously satisfy \(|\gamma_{\mathrm{occ}}|/\pi\ge0.98\) and \(g_{\mathrm{dir}}/t\ge0.05\). Its asymmetric variation with \(\delta_G\) reflects the shifted saddle-shaped gap-closing geometry together with the finite sampling domain and numerical robustness criteria.
    }
    \label{fig:fig6}
\end{figure*}

\section{Ideal-LSS hyperbolic gap boundary and robustness of armchair nanoribbon model}
\label{app:hyperbolic_gap_armchair}

The exact local-support construction is determined by the coupling
vector that connects the parent sector to the auxiliary site. In
particular, when the coupling vector is constructed from the parent
flat-band eigenstates, the local-support block decomposition remains
valid even when the auxiliary site has a nonzero on-site energy
$\varepsilon_G$.  Thus, $\varepsilon_G$ does not by itself destroy the LSS construction.

It can nevertheless shift the energies of the three-level LSS sector.
Consequently, the LSS-induced band may touch a dispersive parent-band edge. The occupied Zak phase can therefore be interpreted as a gapped topological quantity only in the region where the direct occupied gap remains open. We investigated the parameter relation at which this gap closing occurs.

The relevant parent-band edge is
\begin{equation}
\begin{aligned}
E_c
&=
\varepsilon+
\sqrt{\frac{5t^2+t_x^2-\mathcal{R}}{2}},
\\
\mathcal{R}
&=
\sqrt{(3t^2+t_x^2)^2+16t^3t_x}.
\end{aligned}
\label{eq:appC_parent_edge}
\end{equation}

We introduce the dimensionless variables
\begin{align}
e&=\frac{E-\varepsilon}{t},
&
e_c&=\frac{E_c-\varepsilon}{t},
&
\delta_G&=\frac{\varepsilon_G-\varepsilon}{t},
\nonumber\\
\kappa_1&=\frac{g_1}{t},
&
\kappa_2&=\frac{g_2}{t},
&
\tau&=\frac{t_x}{t}.
\label{eq:appC_dimensionless_variables}
\end{align}

The LSS-induced band closes the relevant parent-band gap when its
energy reaches $E_c$.  In terms of the dimensionless variables, this
condition gives
\begin{equation}
\kappa_2^2(1-e_c)
-
\kappa_1^2(1+e_c)
=
(1-e_c^2)(e_c-\delta_G).
\label{eq:appC_hyperbola}
\end{equation}

For fixed $\delta_G$, Eq.~\eqref{eq:appC_hyperbola} is a hyperbola in
the $(\kappa_1,\kappa_2)$ plane.  It is the analytic gap-closing
boundary: crossing this curve changes the direct-gap status of the
occupied subspace.  The boundary is not an additional condition for
the existence of the LSS block decomposition; rather, it specifies
which LSS coupling parameters remain spectrally isolated at the chosen
on-site detuning. Equivalently, solving Eq.~\eqref{eq:appC_hyperbola} for the detuning $\delta_G$
gives the gap-closing surface
\begin{equation}
\delta_G(\kappa_1,\kappa_2)
=
e_c+
\frac{\kappa_1^2}{1-e_c}
-
\frac{\kappa_2^2}{1+e_c},
\qquad e_c\neq\pm1.
\label{eq:appC_gap_closing_surface}
\end{equation}

This is a saddle-shaped surface in the
$(\kappa_1,\kappa_2,\delta_G)$ parameter space.  At
$\kappa_1=\kappa_2=0$, the gap closes at
$\delta_G=e_c$, equivalently at
$\varepsilon_G=E_c$.  Thus, the gap closing starts when the
symmetry-breaking-site energy overlaps the relevant parent-band edge.
Finite $\kappa_1$ and $\kappa_2$ shift this condition according to
Eq.~\eqref{eq:appC_gap_closing_surface}.

For the symmetry-broken model considered here, the occupied-band index is
chosen as $n_{\mathrm{occ}}=4$.
The direct occupied gap is therefore defined by
\begin{equation}
g_{\mathrm{dir}}
=
\min_{k\in\mathrm{BZ}}
\left[E_5(k)-E_4(k)\right].
\label{eq:appC_direct_gap}
\end{equation}

The occupied-subspace Zak phase is evaluated only while
$g_{\mathrm{dir}}>0$.  In the numerical classification used for the
figures, the occupied gap is regarded as open when
\begin{equation}
\frac{g_{\mathrm{dir}}}{t}\geq 0.05.
\label{eq:appC_gap_threshold}
\end{equation}

To quantify the parameter region that simultaneously retains the
occupied phase and an open direct gap, let
\begin{equation}
\mathcal{D}_K=[-K,K]^2
\end{equation}
denote the sampled coupling domain.  The robust fraction is defined as
\begin{align}
f_K(\delta_G)
&=
\frac{1}{|\mathcal{D}_K|}
\int_{\mathcal{D}_K}
d^2\boldsymbol{\kappa}\,
R(\boldsymbol{\kappa};\delta_G),
\label{eq:appC_robust_fraction}
\\
R(\boldsymbol{\kappa};\delta_G)
&=
\mathbf{1}\!\left[
\frac{|\gamma_{\mathrm{occ}}|}{\pi}\geq0.98
\ \wedge\
\frac{g_{\mathrm{dir}}}{t}\geq0.05
\right].
\label{eq:appC_robust_indicator}
\end{align}

On the discrete numerical grid, the integral in
Eq.~\eqref{eq:appC_robust_fraction} is replaced by the corresponding
sample mean.  Therefore, $f_K(\delta_G)$ is a numerical diagnostic
that depends on the chosen coupling range $K$, grid resolution, phase
threshold, and gap threshold; it is not a universal function.

The robust-fraction curve need not satisfy an exact relation of the
form
$f_K(-\delta_G)=1-f_K(\delta_G)$.  The analytic surface is centered at
$\delta_G=e_c$, rather than at $\delta_G=0$, and the two coupling
directions have different curvatures,
$1/(1-e_c)$ and $1/(1+e_c)$.  In addition, the finite sampling domain
and the simultaneous phase and gap thresholds modify the measured
fraction.  The resulting asymmetric decrease of $f_K(\delta_G)$ is
therefore a consequence of the shifted saddle-shaped gap-closing
geometry and the numerical robustness criterion.

\section{Absence of a hidden-symmetry reduction in the minimal two-band construction}
\label{app:no_hidden_symmetry}

In this Appendix, we show that the minimal LSS construction cannot be
reduced to the hidden-symmetry block mechanism underlying the decoupled
topological-subspace construction of Ref.~\cite{xiao2016coexistence}.
To make this distinction precise, we define a hidden-symmetry reduction as
the existence of a $k$-independent unitary transformation that decomposes
the full Bloch Hamiltonian into proper invariant blocks, with the
LSS-protected band contained in a block whose Hamiltonian possesses the
crystalline symmetry responsible for Zak-phase quantization.
Equivalently, there must exist a nontrivial $k$-independent orthogonal
projector $\Pi$, with $0\neq\Pi\neq\mathbbm{1}$, satisfying
\begin{equation}
    [\Pi,H(k)]=0,
    \qquad
    \Pi|U_o(k)\rangle=|U_o(k)\rangle,
    \label{eq:appA_projector_condition}
\end{equation}
for all $k$.
The first condition ensures that the subspace selected by $\Pi$ is invariant
under the full Hamiltonian, while the second requires the entire protected
Bloch band to lie within this fixed subspace.
A hidden-symmetry interpretation would then require the Hamiltonian
restricted to this invariant subspace to retain the crystalline symmetry
that quantizes the Zak phase, even though that symmetry is absent from the
full Hamiltonian.

We emphasize that the $k$ independence of $\Pi$ is essential. For any
nondegenerate Bloch band one can always introduce the spectral projector
$|U_n(k)\rangle\langle U_n(k)|$, or equivalently diagonalize $H(k)$ by a
$k$-dependent unitary matrix. Such a construction exists for every isolated
band and therefore does not constitute a hidden crystalline symmetry.

\subsection{Minimal LSS Hamiltonian}

Consider the two-orbital parent Hamiltonian $h_1(k)$ introduced in the main
text and a single auxiliary orbital,
\begin{equation}
    H(k)=
    \begin{pmatrix}
        h_1(k) & w(k)\\
        w^\dagger(k) & m(k)
    \end{pmatrix},
    \label{eq:appA_fullH}
\end{equation}
where $h_1(k)$ acts in the two-dimensional parent space
$\mathcal H_A\simeq\mathbb C^2$ and $m(k)$ is the auxiliary-sector
Hamiltonian. Let $|u_o(k)\rangle$ and $|u_u(k)\rangle$ denote normalized
occupied and unoccupied eigenvectors of $h_1(k)$, respectively. The LSS
coupling is chosen as
\begin{equation}
    w^\dagger(k)=\lambda(k)\langle u_u(k)|,
    \label{eq:appA_coupling}
\end{equation}
or, equivalently, in terms of the unnormalized eigenvector used in the main
text. Orthogonality immediately gives
\begin{equation}
    w^\dagger(k)|u_o(k)\rangle=0.
    \label{eq:appA_kernel}
\end{equation}
Hence
\begin{equation}
    |U_o(k)\rangle=
    \begin{pmatrix}
        |u_o(k)\rangle\\
        0
    \end{pmatrix}
    \label{eq:appA_protected}
\end{equation}
is an exact eigenstate of the full Hamiltonian.

The question is whether the exact band in Eq.~\eqref{eq:appA_protected}
can be interpreted as belonging to a $k$-independent symmetry-preserving
block of $H(k)$.

\subsection{Nonexistence of a one-dimensional invariant block}

Suppose that a one-dimensional, $k$-independent
invariant subspace $\mathcal{S}$ contains the
protected Bloch state $|U_o(k)\rangle$ for every
$k$. Since $\mathcal{S}$ is one-dimensional,
all protected states must be mutually collinear,
such that
\begin{equation}
|U_o(k)\rangle=e^{i\phi(k)}|U_0\rangle,
\end{equation}
where $|U_0\rangle$ is a fixed normalized vector.

The corresponding spectral projector would then
be independent of momentum,
\begin{equation}
P_o(k)=|U_0\rangle\langle U_0|.
\end{equation}
In the periodic Bloch basis used here, such a
constant projector admits a smooth periodic
eigenvector with a vanishing Berry connection.
Consequently, its Zak phase must satisfy
\begin{equation}
\gamma_o=0\pmod{2\pi}.
\end{equation}
This contradicts the exactly quantized value
$\gamma_o=\pi$ of the protected band.

The same conclusion follows directly from the
inversion eigenvalues of the parent Hamiltonian.
At the inversion-invariant momenta $k_\star=0,\pi$,
the occupied parent states satisfy
\begin{equation}
P|u_o(k_\star)\rangle
=\xi_o(k_\star)|u_o(k_\star)\rangle,
\end{equation}
where $P=\sigma_x$ and $\xi_o(k_\star)=\pm1$.
In the topological regime,
\begin{equation}
\xi_o(0)\xi_o(\pi)=-1.
\end{equation}
Therefore, the two occupied states have opposite
inversion eigenvalues and are orthogonal,
\begin{equation}
\langle u_o(0)|u_o(\pi)\rangle=0.
\end{equation}

It follows that the corresponding embedded states
$|U_o(0)\rangle$ and $|U_o(\pi)\rangle$ are
linearly independent. They cannot both belong
to a one-dimensional, $k$-independent subspace.

Moreover, since the two parent states span the
two-dimensional parent Hilbert space,
\begin{equation}
\operatorname{span}
\left\{
|u_o(0)\rangle,|u_o(\pi)\rangle
\right\}
=\mathcal{H}_A,
\end{equation}
any fixed subspace containing the entire
protected band must contain the full embedded
parent sector. This observation will be used
below to exclude a two-dimensional invariant
block.

\subsection{Equivalent projector formulation}

The same conclusion can be expressed directly in terms of projectors.
Assume that a constant projector $\Pi$ satisfying
Eq.~\eqref{eq:appA_projector_condition} contains the protected band,
\begin{equation}
    \Pi |U_o(k)\rangle=|U_o(k)\rangle
    \qquad
    \text{for all }k.
    \label{eq:appA_contains}
\end{equation}
Because the protected vectors at two momenta span the full embedded parent
space, Eq.~\eqref{eq:appA_contains} implies
\begin{equation}
    \Pi
    \begin{pmatrix}
        |\psi\rangle\\0
    \end{pmatrix}
    =
    \begin{pmatrix}
        |\psi\rangle\\0
    \end{pmatrix}
    \qquad
    \forall |\psi\rangle\in\mathcal H_A.
\end{equation}
Thus, $\operatorname{rank}\Pi\geq 2$. If
$\operatorname{rank}\Pi=2$, its image must coincide with
the embedded parent sector. However, this sector is
not invariant under $H(k)$ because the coupling to
the auxiliary orbital is nonzero. This contradicts
the commutation condition $[\Pi,H(k)]=0$ in
Eq.~(D1). If $\operatorname{rank}\Pi=3$, then
$\Pi=\mathbb{1}$, and the decomposition is trivial.
Therefore, no nontrivial $k$-independent reducing
projector containing the entire protected Bloch
band exists.
If ${\rm rank}\,\Pi=3$, then
$\Pi=\mathbbm{1}$ and the decomposition is trivial. This establishes the
absence of any nontrivial constant reducing projector containing the
protected band.

Notice that this argument uses only three ingredients:
(i) a momentum-dependent protected-band projector,
so that $P_o(k)$ is not constant;
(ii) a single auxiliary degree of freedom; and
(iii) a nonvanishing coupling from the complementary parent direction to
the auxiliary sector. It is therefore more general than the particular
Pythagorean parameterization employed in the main text.

\section{Hidden reflection symmetry of the armchair-nanoribbon construction}
\label{app:armchair_hidden}

The armchair-nanoribbon example has a different mathematical structure.
Here the LSS coupling generates an exact $k$-independent decomposition of
the seven-dimensional Hilbert space into a four-dimensional sector that
retains reflection symmetry and a three-dimensional sector in which the
global reflection can be broken. Thus this particular lattice realization
\emph{can} be recast as a hidden-symmetry construction. We derive this
decomposition explicitly below.

\subsection{Parent Hamiltonian and adapted basis}

In the basis
\begin{equation}
    \mathcal B=(A,B,C,D,E,F),
\end{equation}
the six-site parent Hamiltonian is
\begin{equation}
H_0(k)=
\begin{pmatrix}
\epsilon&t&0&0&0&t\\
t&\epsilon&t&0&t_xe^{-ik}&0\\
0&t&\epsilon&t&0&0\\
0&0&t&\epsilon&t&0\\
0&t_xe^{ik}&0&t&\epsilon&t\\
t&0&0&0&t&\epsilon
\end{pmatrix}.
\label{eq:appB_H0}
\end{equation}
The two $k$-independent flat-band eigenvectors are
\begin{align}
    |v_+\rangle
    &=
    \frac{1}{2}
    (1,0,-1,-1,0,1)^T,
    \\
    |v_-\rangle
    &=
    \frac{1}{2}
    (1,0,-1,1,0,-1)^T .
\end{align}
It is useful to replace them by the equivalent orthonormal combinations
\begin{align}
    |q_1\rangle
    &=
    \frac{|A\rangle-|C\rangle}{\sqrt2},
    &
    |q_2\rangle
    &=
    \frac{|F\rangle-|D\rangle}{\sqrt2},
    \label{eq:appB_qbasis}
\end{align}
for which
\begin{align}
    |v_+\rangle&=\frac{|q_1\rangle+|q_2\rangle}{\sqrt2},
    &
    |v_-\rangle&=\frac{|q_1\rangle-|q_2\rangle}{\sqrt2}.
    \label{eq:appB_vq}
\end{align}

The orthogonal complement of ${\rm span}\{q_1,q_2\}$ in the parent Hilbert
space is spanned by
\begin{align}
    |p_1\rangle&=\frac{|A\rangle+|C\rangle}{\sqrt2},
    &
    |p_2\rangle&=|B\rangle,
    \nonumber\\
    |p_3\rangle&=|E\rangle,
    &
    |p_4\rangle&=\frac{|D\rangle+|F\rangle}{\sqrt2}.
    \label{eq:appB_pbasis}
\end{align}
All six vectors in Eqs.~\eqref{eq:appB_qbasis} and
\eqref{eq:appB_pbasis} are independent of momentum.

Direct application of $H_0(k)$ shows that the two sectors do not mix,
\begin{equation}
    \langle p_i|H_0(k)|q_j\rangle=0
    \qquad
    \forall i=1,\ldots,4,\quad j=1,2.
    \label{eq:appB_parent_decoupling}
\end{equation}
Consequently, in the ordered basis
\begin{equation}
    \mathcal B'
    =
    (p_1,p_2,p_3,p_4;q_1,q_2),
\end{equation}
the parent Hamiltonian is already block diagonal,
\begin{equation}
    U_6^\dagger H_0(k)U_6
    =
    H_4(k)\oplus H_2,
\end{equation}
where
\begin{equation}
H_4(k)=
\begin{pmatrix}
\epsilon & \sqrt2 t & 0 & t\\
\sqrt2 t&\epsilon&t_xe^{-ik}&0\\
0&t_xe^{ik}&\epsilon&\sqrt2 t\\
t&0&\sqrt2 t&\epsilon
\end{pmatrix},
\label{eq:appB_H4}
\end{equation}
and
\begin{equation}
H_2=
\begin{pmatrix}
\epsilon&t\\
t&\epsilon
\end{pmatrix}.
\label{eq:appB_H2}
\end{equation}
The eigenvectors of $H_2$ are precisely $v_\pm$, with eigenvalues
$\epsilon\pm t$.

\subsection{Addition of the auxiliary orbital}

The auxiliary orbital $G$ is coupled to the parent sector through
\begin{equation}
    h_{21}
    =
    \kappa_1 v_+^T+\kappa_2 v_-^T.
    \label{eq:appB_h21}
\end{equation}
Using Eq.~\eqref{eq:appB_vq},
\begin{equation}
    h_{21}
    =
    \frac{\kappa_1+\kappa_2}{\sqrt2}\langle q_1|
    +
    \frac{\kappa_1-\kappa_2}{\sqrt2}\langle q_2|.
    \label{eq:appB_h21_q}
\end{equation}
Hence
\begin{equation}
    h_{21}|p_i\rangle=0
    \qquad
    (i=1,\ldots,4).
    \label{eq:appB_kernel_p}
\end{equation}
This is exactly the destructive-interference condition of the main text,
expressed in the adapted basis.

After adding $G$, choose the seven-dimensional ordered basis
\begin{equation}
    \mathcal B_7'
    =
    (p_1,p_2,p_3,p_4;q_1,q_2,G).
\end{equation}
The corresponding change of basis is $k$ independent. Explicitly, if the
original basis is ordered as $(A,B,C,D,E,F,G)$, one may take
\begin{equation}
U_7=
\begin{pmatrix}
1/\sqrt2&0&0&0& 1/\sqrt2&0&0\\
0&1&0&0&0&0&0\\
1/\sqrt2&0&0&0&-1/\sqrt2&0&0\\
0&0&0&1/\sqrt2&0&-1/\sqrt2&0\\
0&0&1&0&0&0&0\\
0&0&0&1/\sqrt2&0& 1/\sqrt2&0\\
0&0&0&0&0&0&1
\end{pmatrix}.
\label{eq:appB_U7}
\end{equation}
Then the full seven-site Hamiltonian takes the exact block form
\begin{equation}
    U_7^\dagger H_{\rm LSS}(k)U_7
    =
    H_4(k)\oplus H_3,
    \label{eq:appB_block}
\end{equation}
with the same $H_4(k)$ as in Eq.~\eqref{eq:appB_H4} and
\begin{equation}
H_3=
\begin{pmatrix}
\epsilon&t&(\kappa_1+\kappa_2)/\sqrt2\\
t&\epsilon&(\kappa_1-\kappa_2)/\sqrt2\\
(\kappa_1+\kappa_2)/\sqrt2&
(\kappa_1-\kappa_2)/\sqrt2&
\epsilon_G
\end{pmatrix},
\label{eq:appB_H3}
\end{equation}
for real couplings. Complex couplings only require the corresponding
Hermitian conjugates in the last row and column.

Equation~\eqref{eq:appB_block} is stronger than the statement that selected
eigenvectors happen to have zero weight on $G$: it exhibits an exact,
momentum-independent reducing decomposition of the full Hamiltonian.

\subsection{Reflection symmetry of the four-dimensional block}

The parent left--right reflection acts as
\begin{equation}
    A\leftrightarrow F,\qquad
    B\leftrightarrow E,\qquad
    C\leftrightarrow D.
\end{equation}
Within the $p$ sector this becomes
\begin{equation}
    p_1\leftrightarrow p_4,
    \qquad
    p_2\leftrightarrow p_3.
\end{equation}
The corresponding $4\times4$ representation is
\begin{equation}
M_4=
\begin{pmatrix}
0&0&0&1\\
0&0&1&0\\
0&1&0&0\\
1&0&0&0
\end{pmatrix},
\qquad
M_4^2=\mathbbm{1}_4.
\label{eq:appB_M4}
\end{equation}
A direct multiplication using Eq.~\eqref{eq:appB_H4} gives
\begin{equation}
    M_4H_4(k)M_4^{-1}=H_4(-k).
    \label{eq:appB_hidden_reflection}
\end{equation}
Therefore the complete four-dimensional sector containing the relevant
dispersive topological bands possesses an exact reflection symmetry,
although the full seven-site Hamiltonian need not.

At the reflection-invariant momenta $k_\star=0,\pi$, one has
\begin{equation}
    [M_4,H_4(k_\star)]=0,
\end{equation}
so the eigenstates of $H_4(k_\star)$ can be assigned reflection eigenvalues
$\xi_n(k_\star)=\pm1$. For an isolated band or isolated occupied subspace
of this block, the corresponding Zak phase is therefore quantized in the
usual way by the reflection data at $k=0$ and $\pi$. In this sense the
quantized Zak phase of the ideal armchair LSS model admits a conventional
hidden-symmetry interpretation.

\subsection{Where the global reflection is broken}

The $q$ sector transforms differently. Reflection exchanges
\begin{equation}
    q_1\leftrightarrow q_2.
\end{equation}
If the central orbital $G$ is taken to be even under reflection, the
three-dimensional sector would be reflection symmetric only if its two
couplings to $q_1$ and $q_2$ were equal. From
Eq.~\eqref{eq:appB_h21_q}, this condition is
\begin{equation}
    \kappa_1+\kappa_2
    =
    \kappa_1-\kappa_2,
\end{equation}
or
\begin{equation}
    \kappa_2=0.
\end{equation}
For generic $\kappa_2\neq0$, including the asymmetric LSS choices used to
break the lattice reflection, the $H_3$ block does not obey this reflection
constraint. Thus the seven-dimensional Hamiltonian is globally reflection
broken even though $H_4(k)$ remains exactly reflection symmetric.

This result also clarifies the relation between destructive interference
and hidden symmetry in this example. Destructive interference is the
microscopic reason why the $p$ sector belongs to the kernel of the coupling
to $G$, Eq.~\eqref{eq:appB_kernel_p}. Because this kernel is itself a
$k$-independent four-dimensional invariant subspace, the interference
condition produces an exact block decomposition, and the surviving parent
reflection becomes a hidden symmetry of the topological block. Thus, in the armchair model, destructive interference generates an exact
$k$-independent decomposition of the full Hamiltonian into invariant
subspaces. The topological subspace retains the reflection symmetry of the
parent system as a hidden symmetry, which in turn enforces the quantization
of the Zak phase.

This should be contrasted with Appendix~\ref{app:no_hidden_symmetry}. In
the minimal two-band construction the protected states form a
momentum-dependent line bundle whose span over the Brillouin zone exhausts
the parent two-dimensional Hilbert space, while that parent space is not
an invariant subspace of the enlarged Hamiltonian. No proper
$k$-independent topological block therefore exists there. The armchair
model represents a special case in which the LSS kernel happens to define
a fixed invariant subspace and hence admits an additional hidden-symmetry
description.

\bibliography{references.bib}

\end{document}